\documentclass[useAMS,usenatbib]{mn2e}
\usepackage{graphicx}
\title[\textit{Spitzer-IRS} Maps of the Superwind in M82]{Spatially Resolved \textit{Spitzer-IRS} Spectral Maps of the
Superwind in M82}
\author[P. Beir\~ao, L. Armus, M. D. Lehnert, et al.]{P. Beir\~ao$^{1}$\thanks{E-mail: pedro.beirao@obspm.fr}, L. Armus$^{2}$, M. D. Lehnert$^{3}$, P. Guillard$^{3,4}$, T. Heckman$^{5}$, B. Draine$^{6}$, \newauthor
D. Hollenbach$^{7}$, F. Walter$^{8}$, K. Sheth$^{9}$, J. D. Smith$^{10}$, P. Shopbell$^{11}$, F. Boulanger$^{12}$, \newauthor
J. Surace$^{2}$,  C. Hoopes$^{5}$, and C. Engelbracht$^{13}$\\
$^{1}$Observatoire de Paris, LERMA, CNRS, 61 Av. de l'Observatoire, 75014 Paris, France\\
$^{2}$Spitzer Science Center, California Institute of Technology, MC 220-06, Pasadena, CA 91125\\
$^{3}$CNRS, UMR 7095, Institut d'Astrophysique de Paris, 98 bis boulevard Arago, 75014 Paris, France\\
$^{4}$Sorbonne Universit\'es, UPMC Universit{\'e} Paris VI, 4 place Jussieu, 75005 Paris, France\\
$^{5}$Center for Astrophysical Sciences, Department of Physics and Astronomy, Johns Hopkins University, Baltimore, MD 21218, USA\\
$^{6}$Princeton University Observatory, Peyton Hall, Princeton, NJ 08544-1001, USA\\
$^{7}$SETI Institute, Mountain View, CA 94043, USA\\
$^{8}$Max-Planck Institut f\"ur Astronomie, K\"onigstuhl 17, D-69117 Heidelberg, Germany\\
$^{9}$North American ALMA Science Center, National Radio Astronomy Observatory, 520 Edgemont Road, Charlottesville, VA 22901, USA\\
$^{10}$Ritter Astrophysical Observatory, University of Toledo, Toledo, OH 43606, USA\\
$^{11}$Astronomy Department, California Institute of Technology, MC 249-17, Pasadena, CA 91125\\
$^{12}$Institut d'Astrophysique Spatiale (IAS), UMR 8617, CNRS \& Universit\'e Paris-Sud 11, B\^atiment 121, 91405, Orsay Cedex, France\\
$^{13}$Steward Observatory, University of Arizona, 933 North Cherry Avenue, Tucson, AZ 85721, USA
}

\begin{document}
 
\date{}

\pagerange{\pageref{firstpage}--\pageref{lastpage}} \pubyear{2014}

\maketitle

\label{firstpage}

\begin{abstract}

We have mapped the superwind/halo region of the nearby starburst galaxy M82 in the mid-infrared with
$Spitzer-IRS$. The spectral regions covered include
the H$_2 S(1)-S(3)$, [NeII], [NeIII] emission lines and PAH features.
We estimate the total warm H$_2$ mass
and the kinetic energy of the outflowing warm molecular gas to be between $M_{warm}\sim5-17\times10^6$ M$_{\odot}$ and $E_{K}\sim6-20\times10^{53}$ erg.
Using the ratios of the 6.2, 7.7 and 11.3 micron PAH features in the IRS
spectra, we are able to estimate the average size and ionization state
of the small grains in the superwind. There are large variations in the PAH
flux ratios throughout the outflow. The 11.3/7.7 and the
6.2/7.7 PAH ratios both vary by more than a factor of five across the
wind region.  
The Northern part of the wind has a
significant population of PAH's with smaller 6.2/7.7 ratios than either
the starburst disk or the Southern wind, indicating that on average, PAH emitters are larger and more ionized.
The warm molecular gas to PAH flux ratios (H$_2/PAH$) are enhanced in
the outflow by factors of 10-100 as compared to the starburst disk. This enhancement in the H$_2/PAH$ ratio does not seem to
follow the ionization of the atomic gas (as measured
with the [NeIII]/[NeII] line flux ratio) in the outflow. This suggests that much of the warm H$_2$ in the outflow is excited by shocks. 
The observed H$_2$ line intensities can be reproduced with low velocity shocks ($v < 40$ km s$^{-1}$) driven into moderately dense molecular gas ($10^2 <n_H < 10^4$
cm$^{-3}$) entrained in the outflow. 

\end{abstract}

\begin{keywords}
galaxies: starburst --- galaxies: individual (M82) --- infrared: galaxies
\end{keywords}

\section{Introduction}

Galactic outflows, ``superwinds'', are driven by hot gas and momentum
generated by the combined mechanical and radiative energy and momentum
output from stellar winds and supernovae \citep[e.g.][]{heckman90}. They
have been invoked as a source for the heating and metal-enrichment of
both the intra-cluster and inter-galactic medium \citep{adel03,loewe04},
as a major driver of galaxy evolution \citep[e.g.][]{croton06},
and as playing a crucial role in establishing the mass-metallicity
relation among galaxies \citep{tremonti04}. Superwinds have also been
hypothesized to be the mechanism through which a dust enshrouded
ultraluminous infrared galaxy (ULIRG) transitions to an optically
identified quasar \citep[e.g.][]{sanders88}.

The dynamical evolution of a starburst-driven outflow has been extensively
discussed \citep[e.g.][]{chevalier85, suchkov94, wang95, tenorio98,
strick00}. Superwinds are generated when the kinetic energy in the outflow
from massive stars and supernovae is thermalized, generating a region of
very hot ($T\sim10^8$ K) low-density gas in the ISM of a starburst galaxy
\citep{chevalier85}. This hot gas will expand, sweeping-up and entraining
ambient gas and creating a hot cavity within the overall structure of the
ISM. This region of hot gas is highly over-pressurized relative to its
surrounding ambient ISM and will expand along the steepest pressure gradients.  As it
expands, it sweeps up and shocks ambient material, creating an outflowing
``superwind" \citep{castor75, weaver77}. Contributing to this overall
energy driven outflow is the radiation pressure on dust grains that are
likely coupled to the ionized gas within the flow \citep{murray05}.

There is a significant amount of morphological, physical and kinematic
evidence for the existence of superwinds in nearby starburst and
infrared luminous galaxies \citep[e.g.][]{heckman90}. Large- scale
optical emission-line and associated X-ray nebulae are ubiquitous in
starbursts \citep{heckman90, lehnert99, strick04}, and these increase
in size and luminosity from dwarf starbursts to ULIRGs,
where they can be tens of kpc in size \citep{grimes05}. In addition,
the winds are observed in molecular lines such as CO \citep{walter02,bolatto13}
and blue-shifted absorption features (e.g. NaD) seen projected against the starburst
nuclei \citep{heckman00, rupke05, martin05} indicating that the
neutral, cold ISM is being swept-up in the outflow. Recently, vigorous
outflows have also been detected in ULIRGs and LIRGs in molecular absorption
lines with $Herschel-PACS$ which suggest very high mass outflow rates
\citep[e.g.][]{sturm11, fischer10, veilleux13, spoon13}. 
ULIRGs which have higher AGN bolometric fractions appear to have higher terminal velocities and shorter gas depletion timescales than those with lower AGN bolometric fractions, suggesting that at least in ULIRGs, the AGN power is related to the feedback on the cold molecular gas \citep{veilleux13,cicone13}.

M82 is the  brightest, nearest, and best-studied starburst galaxy with
an outflow \citep{heckman90, shopbell98, lehnert99}. At a distance of
about 3.5 Mpc \citep{jacobs09}, M82 affords an unparalleled opportunity to study a dusty
outflow in great detail. The outflow in M82 has extensive and complex
extra-planar polycyclic aromatic hydrocarbon (PAH) and H$_2$ emission,
evident from the $Spitzer -$ IRAC and $Spitzer-$IRS observations presented in \citet{engel06}.  Extensive PAH emission is also seen at $3.3\mu$m in the AKARI data \citet{yamagishi12}. PAHs are thought to be the origin
of strong, broad emission features arising in the mid-infrared spectra
of star forming galaxies \citep[e.g.][]{brandl06,smith07a}. M82 has also been
observed in the [CII] and [OI] emission lines by \citet{contursi13}
who find a low velocity outflow in the atomic gas. Complex
extra-planar warm H$_2$ knots and filaments extending more than $\sim5$
kpc above and below the galactic plane of M82 have been imaged in the near-infrared using narrow-band filters by \citet{veilleux09}. Using wide-field imaging, they found that H$_2$ is widespread, suppressed relative to PAHs in the fainter regions, and concluded that H$_2$ is not a dynamically important component of the outflow. However, \citet{veilleux09} could only measure the very hot ($\sim 1000$ K) H$_2$ gas, and the mass of very hot H$_2$ gas is much smaller than the mass of the H$_2$ gas probed by this study, as we show in Section 3.3.

In this paper, we study the M82 outflow with $Spitzer-IRS$. With Spitzer/IRS spectroscopy, we can study in detail all the PAH features, derive the grain properties through analysis of the PAH ratios, and compare these to the warm (200K) molecular gas and the ionized gas (via the strong Ne lines) as a function of position above and below the plane of M82 for the first time. The rotational lines of H$_2$, observable in the infrared, unlike the near-infrared ro-vibrational lines, probe most of the warm molecular gas mass. We then use the ratio of pure rotational H$_2$ lines to constrain the physical and excitation conditions of the gas.
We use PAH 11.3/7.7 and 6.2/7.7 ratio maps to study PAH sizes and ionization across the outflow, [NeIII]/[NeII] to study the variation of the hardness of the ionization field, H$_2/PAH$ to diagnose the origin of the H$_2$ emission, and the H$_2$/[NeII] and PAH/[NeII] line flux ratios to help understand the survivability of the PAH grains in the wind.
For the first time we can make robust
maps of these ratios throughout the outflow of M82,
and compare all of these to the spatially-resolved X-ray emission to
understand the properties of the dust and molecular gas and ionized gas in
the wind. 

The paper is organized as follows: in \S~\ref{obs} we describe
the observations and the data reduction; in \S~\ref{results} we present
and analyze the the spectral maps built from the observations, using the PAH 11.3/7.7 and 6.2/7.7, [NeIII]/[NeII], H$_2/PAH$, H$_2$/[NeII], and PAH/[NeII] line ratios, and also H$_2$ excitation diagrams
to diagnose the properties of the ionized gas, PAHs and the H$_2$ emission; in \S~\ref{discussion}
we diagnose the origin of H$_2$ emission using X-ray images and shock
excitation models; and in \S~\ref{conclusion} we present the conclusions.

\section{Observations and Data Reduction}
\label{obs}

\begin{figure}
\centering
\includegraphics[width=0.5\textwidth]{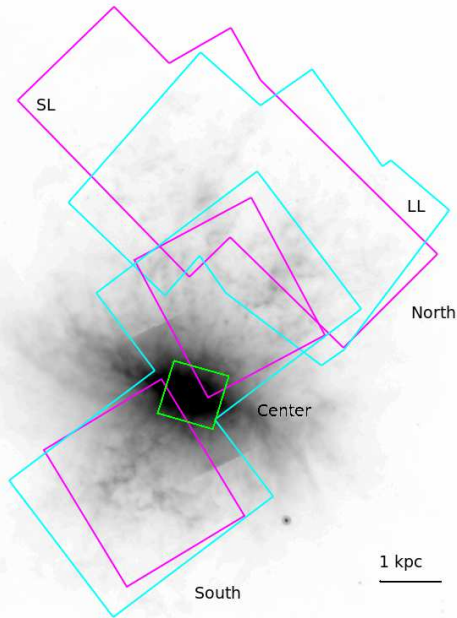}
\caption{Overlay of the footprints of SL ($5-14\mu$m -- magenta) and LL ($14-35\mu$m -- cyan) maps on a $Spitzer-$IRAC 8$\mu$m image \citep{engel06}. The ``center'' region of M82 mapped in \citet{beirao08} is presented in green. The scale is 1 kpc per arcminute.}
\label{footprintsfig}
\end{figure}

\begin{figure}
\centering
\includegraphics[width=0.5\textwidth]{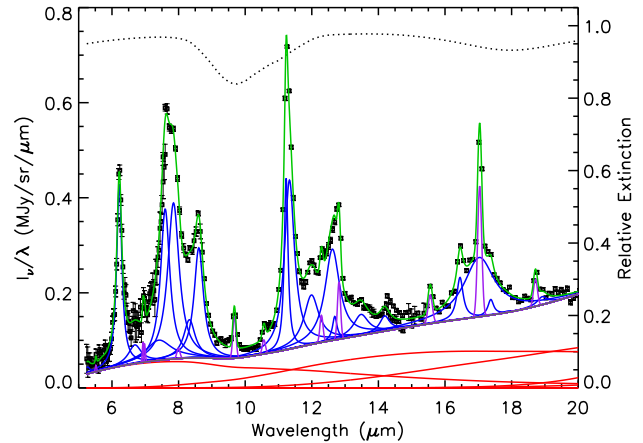}
\caption{Detailed PAHFIT spectral decomposition of a $5 - 20\mu$m spectrum of the M82 outflow. The red solid lines represent the thermal dust continuum components, and the thick gray line shows the total (dust+stellar) continuum. Blue peaks set above the total continuum are PAH features, while the violet peaks are unresolved atomic and molecular spectral lines. A fully mixed dust extinction model is used (shape indicated by the dotted black line at the top), with the relative extinction given on the axis at right. The solid green line is the full fitted model, overlaid on the observed flux intensities and uncertainties.}
\label{pahfitfig}
\end{figure}

The outflow of M82 was observed with $Spitzer$-IRS between November
10--14 2006 and May 9--22 2008, using all low-resolution modules. In
Figure~\ref{footprintsfig} we show the footprints of the IRS observations
on the outflow of M82. We extracted only the maps that contain
both modules SL1 ($7.8 - 14\mu$m) and SL2 ($5.5 - 7.8\mu$m) or LL1 ($20 - 35\mu$m) and LL2 ($14 - 20\mu$m). In total we mapped an area of approximately 5 arcmin$^2$ above and below the plane of M82.

We used CUBISM \citep{smith07a} to build SL and LL spectral cubes from
from the Spitzer-IRS Basic Calibration Data (BCDs). In order to create and subtract a local background from the wind spectra, we observed a region 20 kpc North of the M82 disk with both IRS modules. This region was observed with the same parameters (integration time per pixel and cycles per position) that were used for the wind.  After verifying that no line, PAH or dust continuum was visible in the off field, we subtracted an average image created from this region from the wind data at each location and for each IRS module before building the spectral cubes.

We used PAHFIT \citep{smith07b} to create maps of spectral
lines and PAH features. PAHFIT is a
spectral fitting routine that decomposes IRS low-resolution spectra into
broad PAH features, unresolved line emission, and continuum emission from
dust grains. PAHFIT allows for the deblending of overlapping features,
in particular the PAH emission, silicate absorption and fine-structure atomic, and warm H$_2$ emission lines. 
Because it performs a simultaneous fit to the emission features
and the underlying continuum, PAHFIT works best if it can fit combined
SL and LL spectra across the full $5-40\mu$m range. In order to provide a stable long wavelength continuum from
which to estimate the true PAH emission, including the broad feature
wings, we formed a combined SL + LL cube. In order to investigate the spatial variations of the H$_2/PAH$ ratio at the maximum resolution we decided to rotate and regrid
the LL cubes to match the orientation and pixel size of the SL cubes
($1\arcsec.85\times1\arcsec.85$). Thus we provide a stable long-term extrapolation of the LL fit to the SL fit.
By calculating the flux in the new LL
cubes using a bilinear interpolation (in surface brightness units), we
assure that the flux over the native SL pixel scale is conserved. 

\begin{figure*}
\centering
\includegraphics[width=0.45\textwidth]{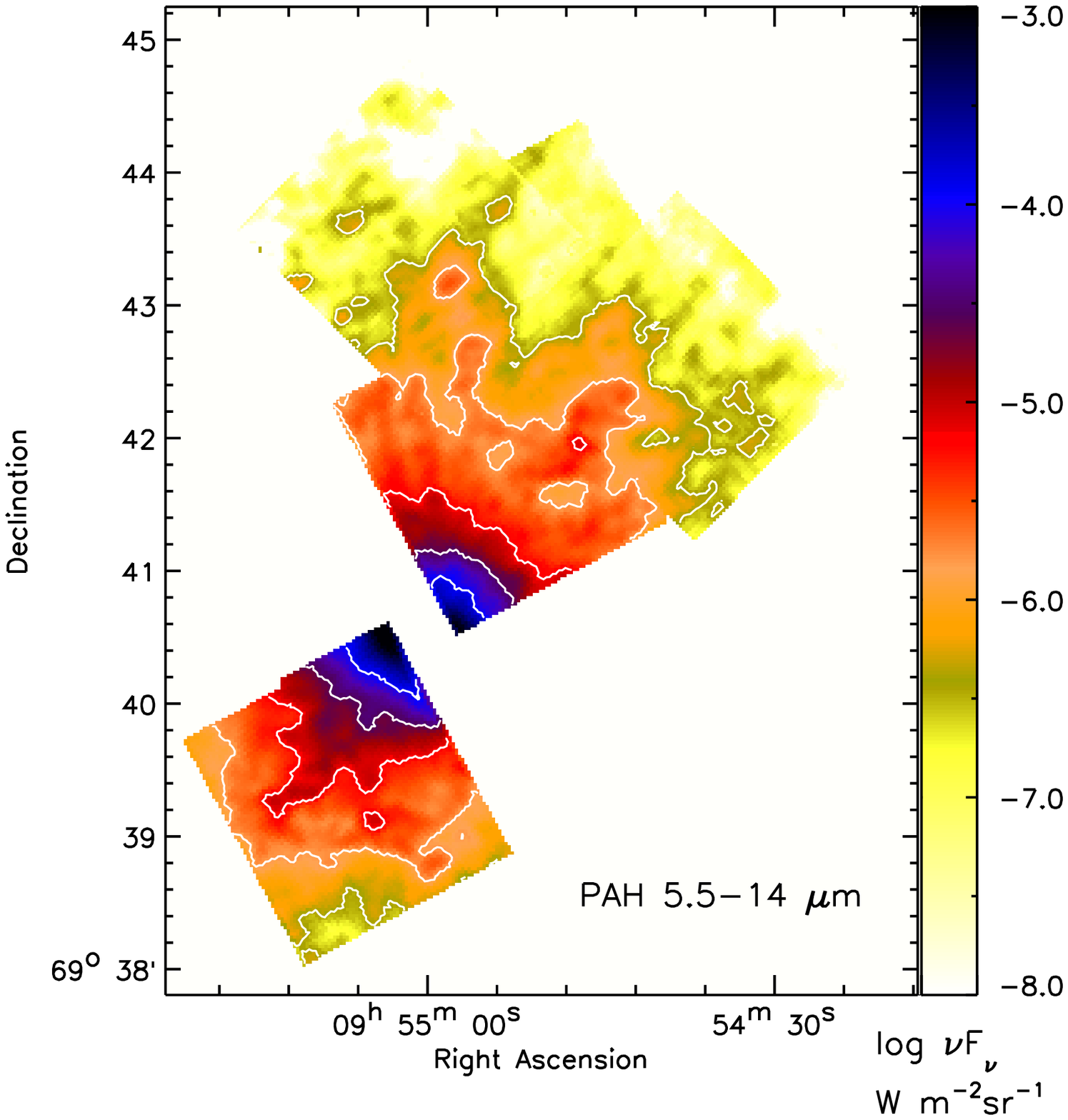}
\quad
\includegraphics[width=0.45\textwidth]{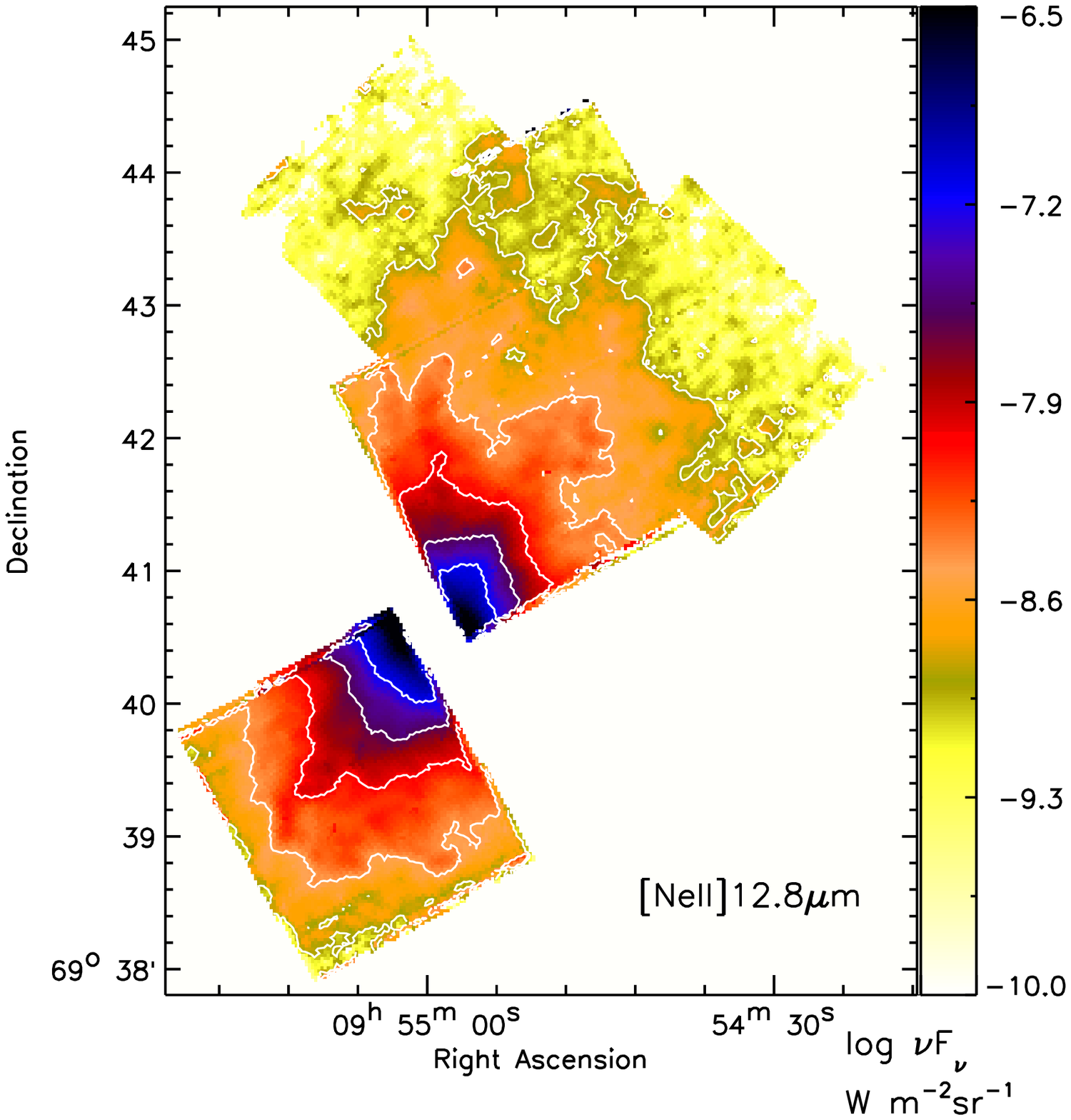}
\quad
\includegraphics[width=0.45\textwidth]{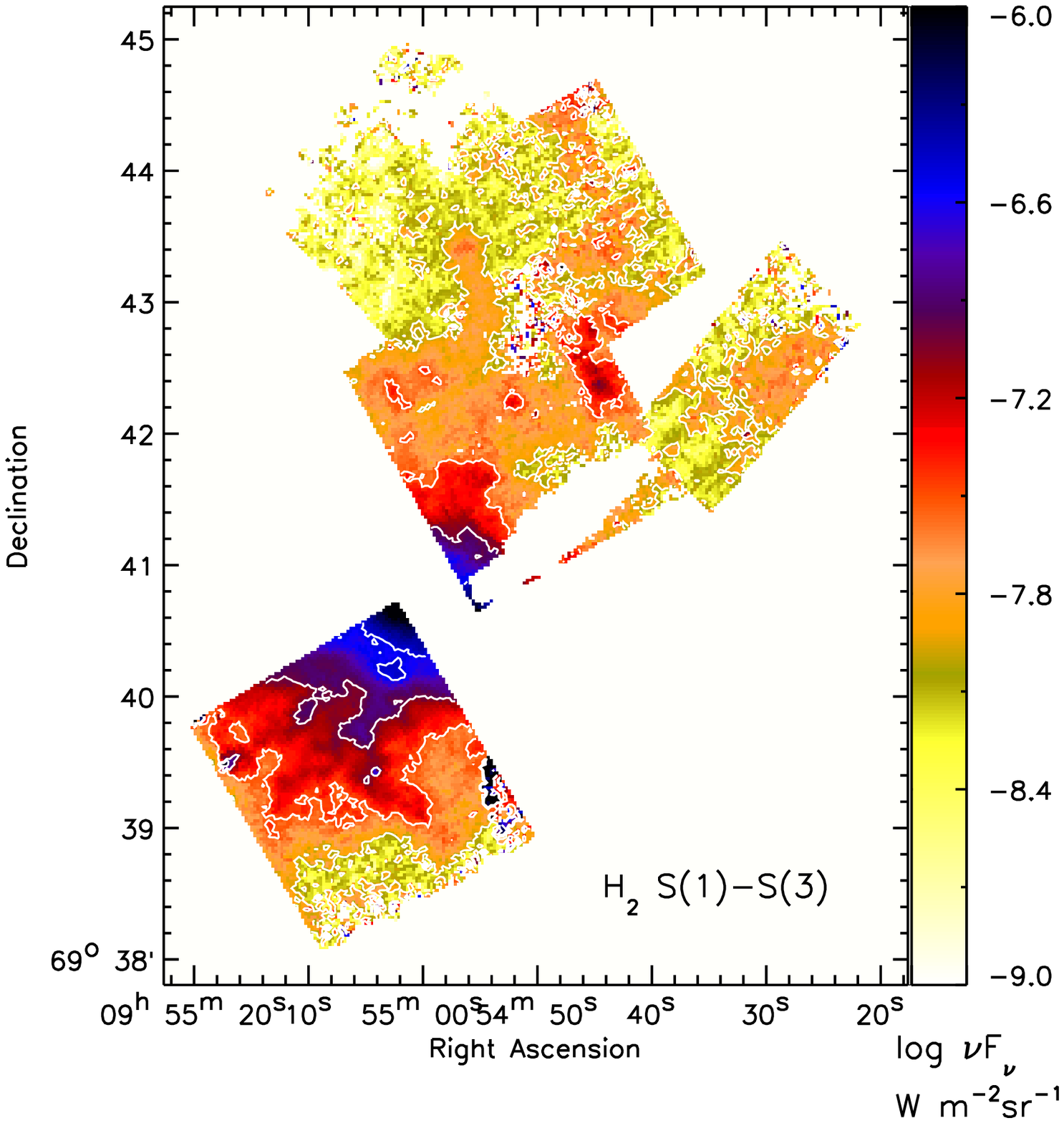}
\quad
\includegraphics[width=0.45\textwidth]{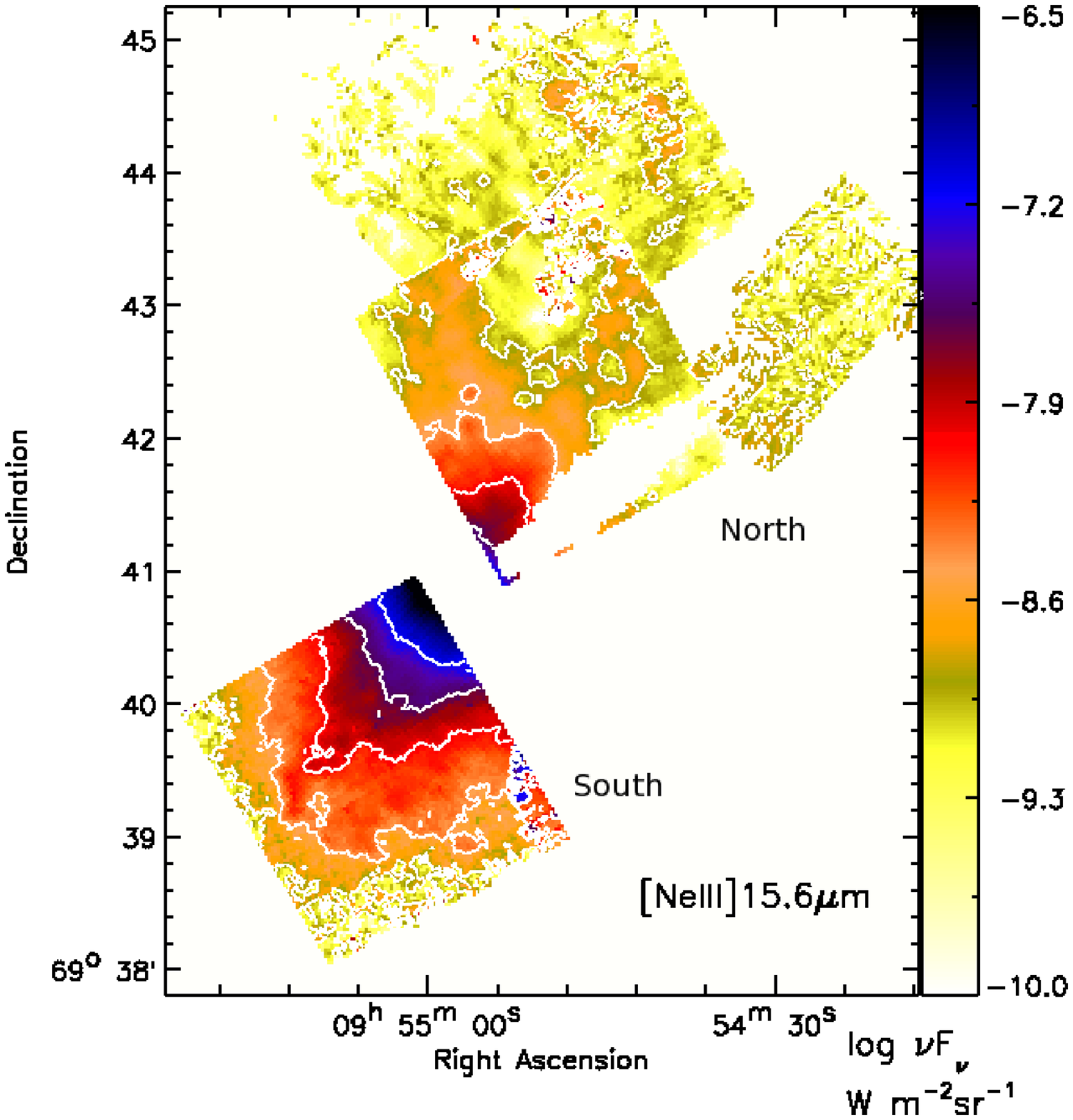}
\quad
\caption{Maps of the continuum-subtracted total PAH emission
between $5-14\mu$m (upper left), the [NeII]$12.8\mu$m
emission (upper right), total H$_2$ S(1)-S(3) emission (lower left)
and total [NeIII]$15.6\mu$m (lower right). The maps are clipped at
2-sigma level in all cases. 
The white radial strip in the bottom two
panels are regions that have been masked out due to scattered light
in the LL2 module, which affects the H$_2$S(1) and [NeIII]$15.6\mu$m
line maps.}
\label{pahmapfig}
\end{figure*}

Finally, we applied
a small correction factor to each individual SL spectrum, necessary to match the continuum in the corresponding LL spectrum. This scale factor ranges between $0.65-0.99$, and was calculated by comparing pixels in the spectral overlap region between SL and LL at approximately
$14\mu$m. This scale factor is applied before running PAHFIT on the
combined SL + LL cubes. Such scaling is commonly observed in
spatially extended sources as a result of the different entrance
slit dimensions of the SL and LL spectrograph modules. In Figure~\ref{pahfitfig} we present an example of the spectral decomposition performed at every location of the map with PAHFIT. The output of PAHFIT is saved for each feature, and is used to create all the maps (PAH, H$_2$, [NeII] and [NeIII]) at
the native resolution of the SL data as shown in Figure~\ref{pahmapfig}.
Although care was taken to avoid the FIR-bright nucleus of M82 during the
spectral mapping, the LL1 data exhibit a ``ripple'' pattern that varies
spatially and spectrally, and the LL2 data have a slightly elevated
continuum level in some pixels (compared to SL1).  We believe both of
these to be the effect of light scattered into the nuclear IRS entrance slits. Therefore
we have opted not to use the LL1 data in the fitting and analysis,
except for the derivation of an upper limit on the H$_2$ S(0) line flux, calculated as the $3-\sigma$ noise level at $28\mu$m.
We have excluded regions of the LL2 data
cubes where the implied scale factor in the continuum between SL1 and
LL2 is greater than 2.0.

\section{Results}
\label{results}

\subsection{Morphology of the dust, ionized gas and warm H$_2$ emission}

\begin{figure*}
\centering
\includegraphics[width=0.45\textwidth]{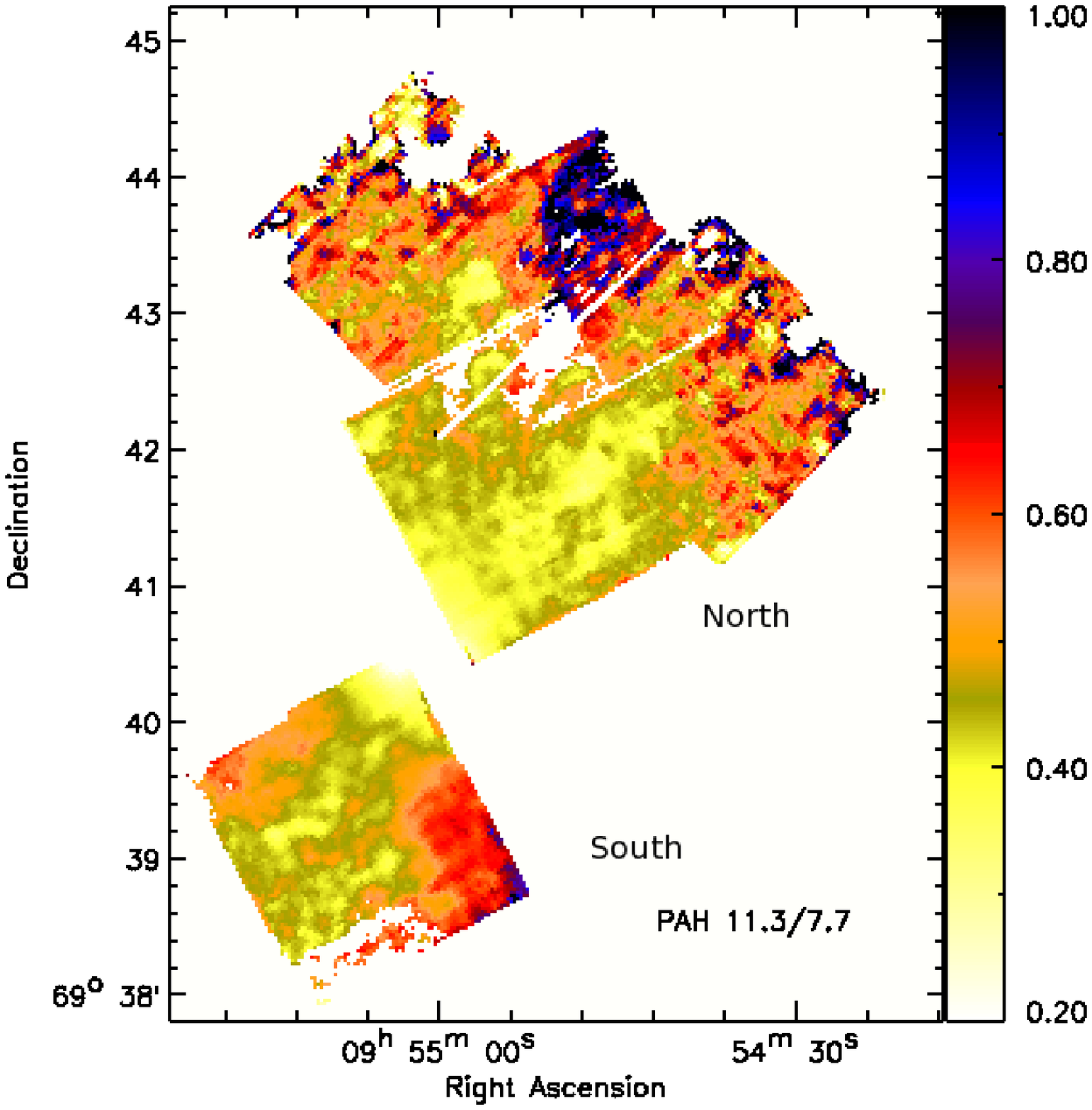}
\includegraphics[width=0.45\textwidth]{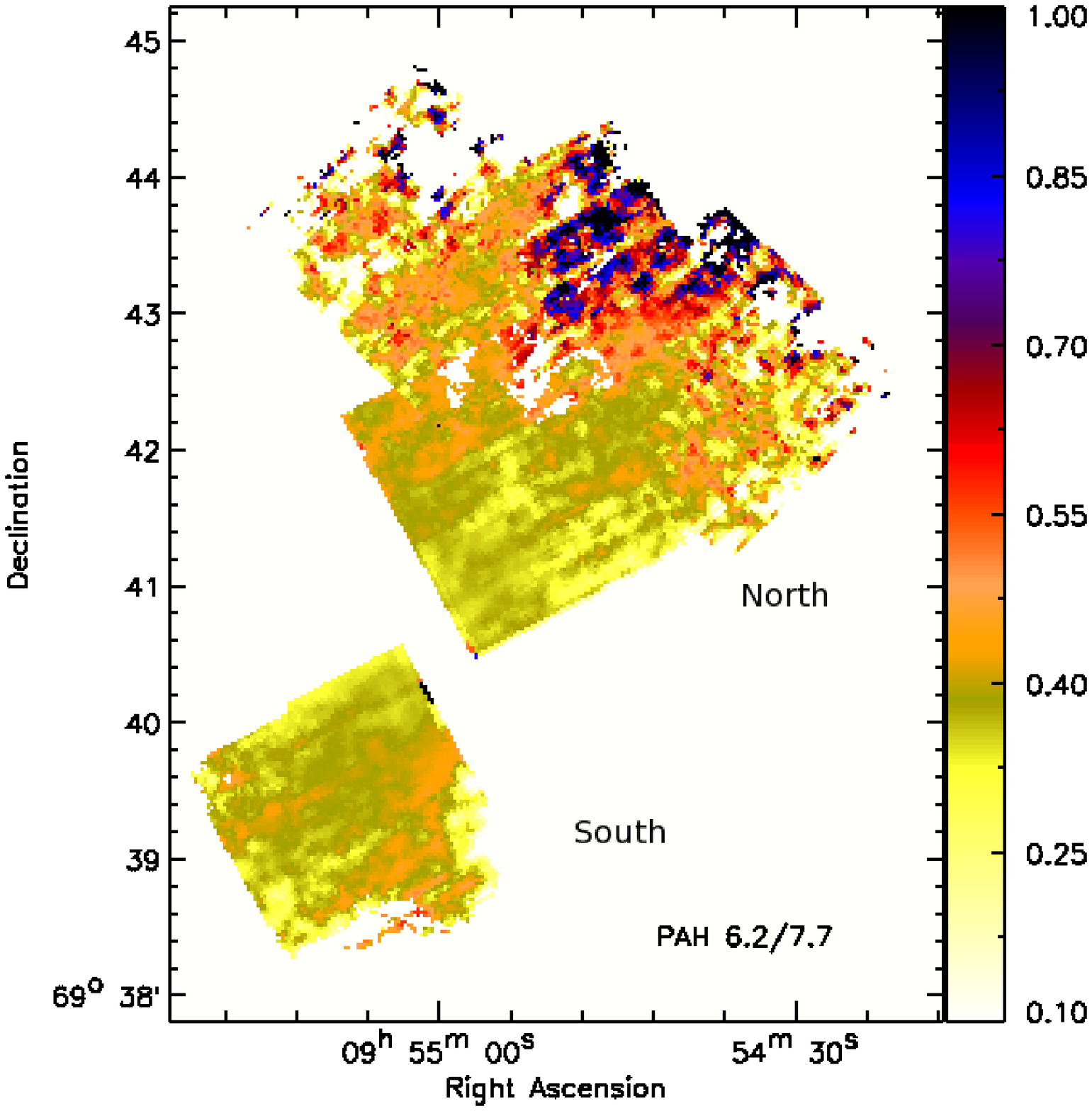}
\includegraphics[width=0.45\textwidth]{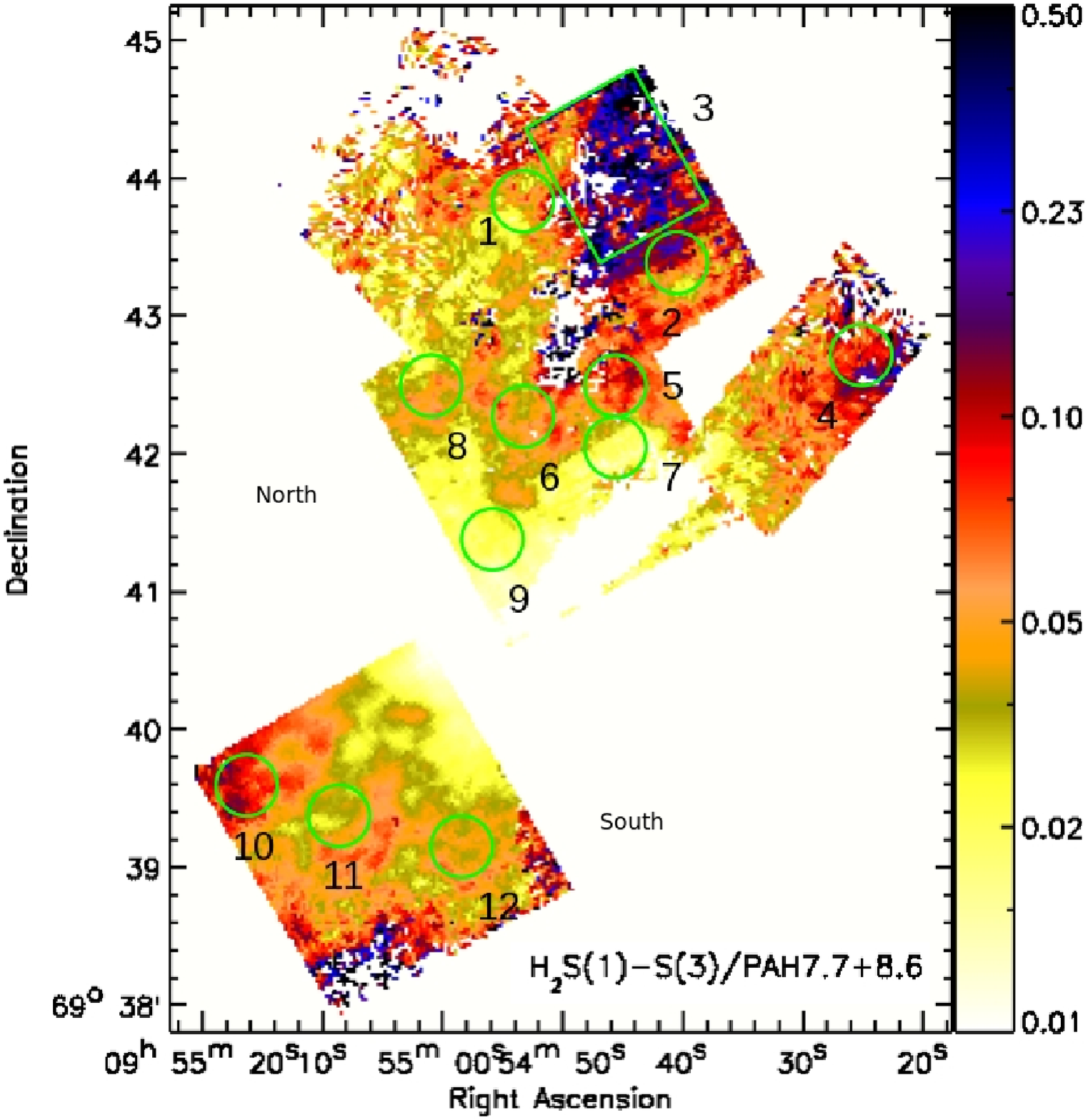}
\includegraphics[width=0.45\textwidth]{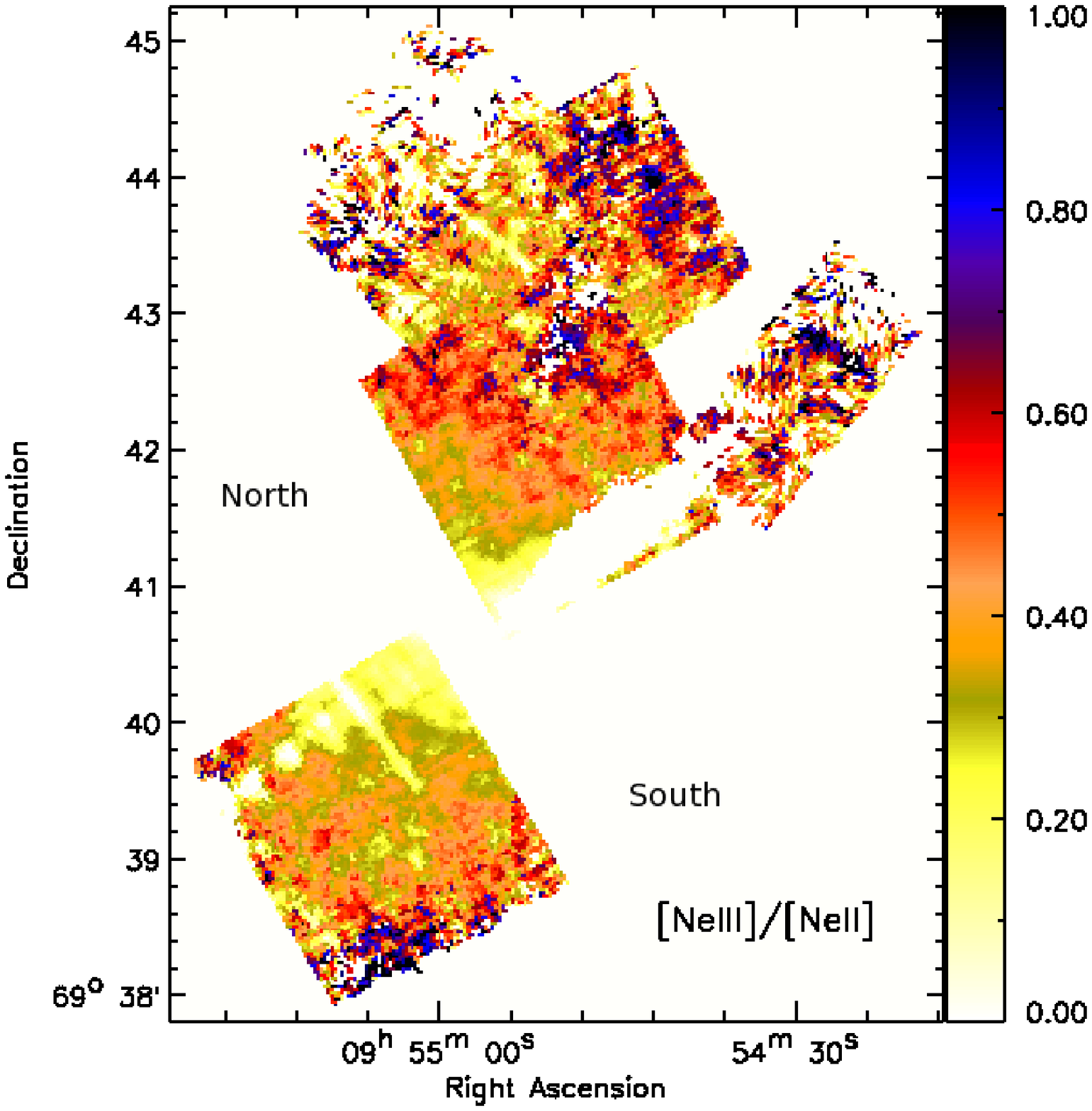}
\caption{Maps of the PAH 11.3/7.7 ratio (upper left), PAH 6.2/7.7 (upper
right), H$_2 [S(1)-S(3)]$/PAH($7.7+8.6\mu$m) (lower left) and [NeIII]/[NeII]
(lower right). All maps have been clipped at the $3-\sigma$ surface
brightness level. The white radial strip in the bottom two panels are
regions that have been masked out due to scattered light in the LL2
module, which affects the H$_2 S(1)$ and [NeIII] lines. The circles
represent the selected regions where the spectra in \S~\ref{spectrumfig}
were extracted and analyzed. The fluxes are presented in Table 1 and 2
(see \S 2 for details).}
\label{ratiomapsfig}
\end{figure*}

In Figure~\ref{pahmapfig} we present the total PAH emission map
calculated by summing over all the PAH bands between $5-14\mu$m, the
[NeII]12.8$\mu$m flux map derived from the SL data cubes, the total H$_2$
map derived from the SL and LL2 cubes, and the [NeIII]15.6$\mu$m emission
map. We find widespread emission with significant structure in all of the line maps.  To the North, both PAH and [NeII] exhibit two prominent, broad radial filaments.  To the South, the PAH map shows a complex structure with a bubble, or loop. This loop is also seen in the [NeII] map.
The overall morphologies of the warm H$_2$ and
H$\alpha$ are generally very similar, as seen in the NIR by  \citet{veilleux09}. Maps of the cold molecular gas made in CO (J=1-0) by \citet{salak13} are at lower spatial resolution, but also show extended, extraplanar gas.

We can study the physical properties of the ionized gas and PAHs by
mapping the ratios between ionized lines such as [NeIII]/[NeII] and
between the different PAH bands. If the ionized gas is photoionized,
the Ne line ratio is sensitive to the hardness of the radiation field
and the ionization parameter. The Ne ratios observed in the outflow are standard values for the hardness of the radiation field found in starburst galaxies. The ratio of the 6.2, 7.7 and 11.3 micron
PAH features can be used to estimate the ionization state and the size
of the small dust grains \citep[e.g.][]{draine01}. Even the overall absence of
PAH emission may be related to the ionization state of the gas since we do not expect gas emitting strongly in [NeIII] to show much PAH emission, as the small grains are easily destroyed by energetic photons \citep[e.g.][]{engel06,madden06,wu06}. The 11.3$\mu$m/7.7$\mu$m ratio is high when PAHs are more
neutral and larger, while the 6.2$\mu$m/7.7$\mu$m PAH ratio is high when PAH
grains are smaller. However, we caution that variations
in the PAH line ratios are also influenced by the structural properties
and elemental abundances in small particles giving rise to the PAH features \citep[e.g.][]{yamagishi12}.

In Figure~\ref{ratiomapsfig} we present maps of the PAH 11.3/7.7 and
6.2/7.7 ratios, the H$_2/PAH$ ratio and the [NeIII]/[NeII]
ratio. The H$_2/PAH$ ratio map was produced using a H$_2$
[S(1)-S(3)] total flux map and PAH ($7.7+8.6\mu$m) flux map.  Although there
is a great deal of structure in the PAH ratio maps, the general trend
is that the 11.3/7.7 and the 6.2/7.7 ratios increase as we move away from the starburst disk. 
The PAH 11.3/7.7 is enhanced by a factor of $\sim 4$
and the PAH 6.2/7.7 is enhanced by a factor of $\sim 3$ at the northern
edge of the outflow compared to the regions near the center of the starburst disk. There is
an enhancement of the 11.3/7.7 and 6.2/7.7 ratios by $\sim 2$ in the
southeast edges of the outflow.

The H$_2/PAH$ ratio
varies between $0.01 - 1$, and the [NeIII]/[NeII] ratio varies
between $0.2-0.9$.  
In the region between the
PAH emission ``fingers'' (as particularly evident in the Northen outflow region in Figure~\ref{pahmapfig}), both maps show an enhancement of the ratio.

\begin{figure}
\includegraphics[width=0.45\textwidth]{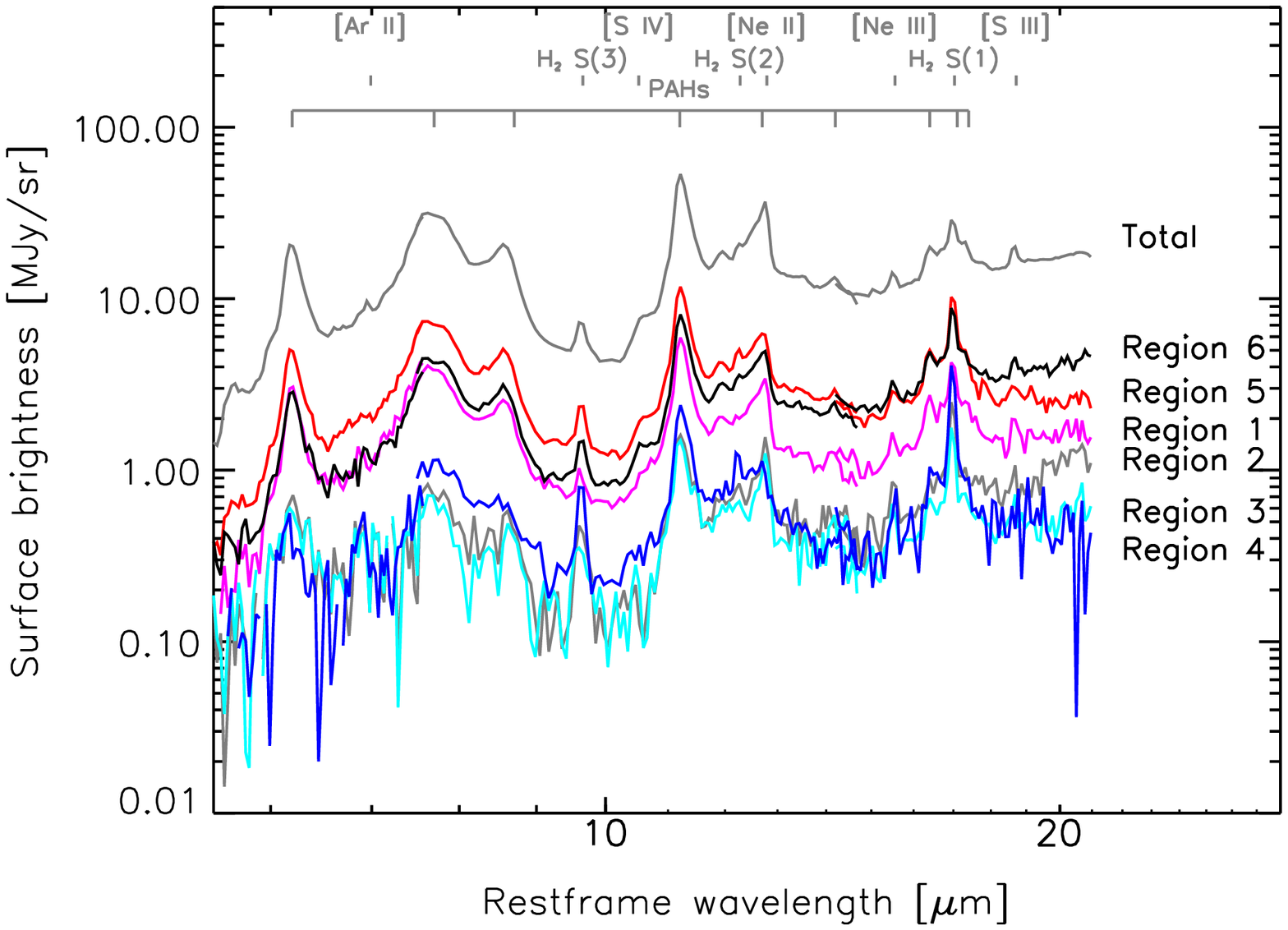}
\includegraphics[width=0.45\textwidth]{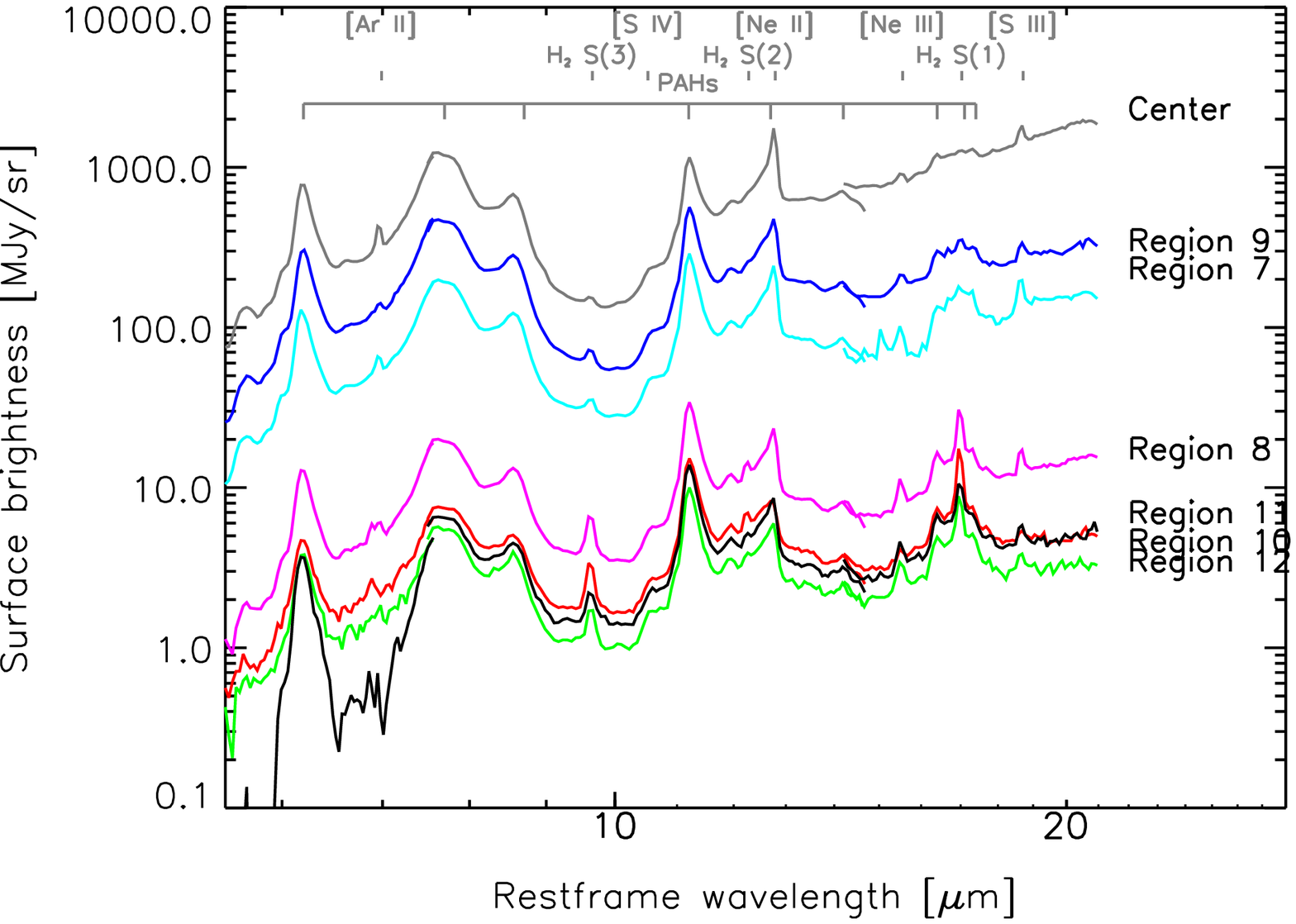}
\caption{Mid-infrared $5-20\mu$m spectra from selected regions in the
outflow, as seen in Figure~\ref{pahmapfig}, and the center of M82 from
\citet{beirao08}. The spectrum labeled ``Total'' is the sum of all pixels in the map.}
\label{spectrumfig}
\end{figure}

In Figure~\ref{ratiomapsfig} we also mark the 12 regions from which we have extracted $5-20\mu$m spectra and performed multi-temperature fits to the H$_2$ lines.  We show these spectra in Figure~\ref{spectrumfig}. These regions were chosen to represent
the full range of H$_2/PAH$ ratios and to approximately sample the
full spatial extent of the outflow. For comparison, we also show an
integrated spectrum of the center of M82 \citep[Figure~\ref{spectrumfig};
][]{beirao08}, which we use as a reference. In addition to the 12 regions, we also show a composite spectrum of the entire IRS map in Figure~\ref{spectrumfig}. We measure the fluxes of the PAH features between $5-14\mu$m, the fine-structure lines [NeII]
and [NeIII], and the rotational excitation lines of H$_2$ from S(0)
to S(3) and present the results in Tables 1 and 2.  Regions 1, 5, 6,
7 and 9 are the most similar to the reference spectrum in slope and the
relative strength of PAH features, which is not surprising since they are
located close to the center. PAHs are also strong in regions 3, 8, 10,
11 and 12, but their spectrum also exhibits a flatter continuum, probably
indicating the presence of colder dust. Region 3 is the most distant from the central starburst and it has the weakest PAH emission. Region 4 exhibits a flat continuum, but a comparatively strong PAH $11.3\mu$m
feature. 

As we have seen in the maps, the spectra show strong variations of the H$_2/PAH$ ratios between
the regions. The H$_2 S(1)$ line is very prominent in the spectrum
of Region 3 compared to the PAH bands, while in the regions close to
the disk, such as Regions 7 and 9, this line is much fainter compared
to the PAH features. Indeed, using the values in Tables 1 and 2, we
calculate in Region 3 a very high H$_2/PAH\sim0.08$, while in Region
9, H$_2/PAH\sim0.01$.

\begin{figure}
\includegraphics[width=0.45\textwidth]{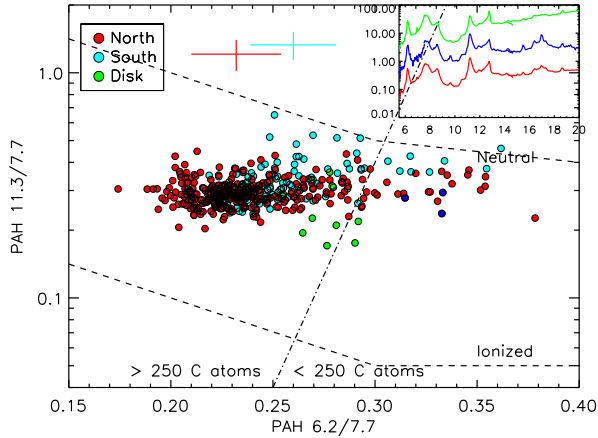}
\caption{Plot of the PAH ratio 11.3/7.7 as a function of 6.2/7.7 for $10\arcsec\times10\arcsec$ regions in the outflow. Red represents the northern outflow, cyan the southern outflow and green represents the central 1 kpc in M82 from \citet{beirao08}. The dashed lines represent neutral and ionized PAHs from \citet{draine01}. The dot-dashed line represent the PAH ratio values for a PAH grain of 250 carbon atoms. The red and cyan crosses represent typical uncertainties in the north and southern outflows.
The three spectra on the upper right corner are derived from different regions in the outflow (in units of MJy/sr). The blue spectrum is derived from a region 2 kpc north of the disk and is represented by the blue data points, the red spectrum is an average spectrum of the whole northern outflow, and the green spectrum the average spectrum of the central region.}
\label{drainefig}
\end{figure}

In Figure~\ref{drainefig} we present a plot of the PAH 11.3/7.7 ratio as
a function of the 6.2/7.7 PAH ratio. The range of 6.2/7.7 PAH ratios in
the outflow is much larger and the values are, on average, smaller than the
disk. The outflow exhibits 6.2/7.7 and 11.3/7.7 ratios ranging between $\sim0.2 - 0.4$. On average, the bulk of the superwind shows larger, more ionized grains than in the starburst disk, although there are regions where the ratios overlap. 
As an example of the different PAH ratios found in the outflow, we present in the upper right corner of Figure~\ref{drainefig} two spectra from different regions of the outflow, and an average spectrum of the disk. In general, the outflow has stronger $17\mu$m PAH and H$_2$ emission than the SB disk, and the spectrum in the region 2 kpc north of the disk (the blue spectrum shown in Figure~\ref{drainefig}) has higher $17\mu$m feature and H$_2$ S(1) flux, indicating the prevalence of larger PAH grains and enhanced H$_2$ emission further away from the disk.
In summary, we see a much larger range in the PAH ratios in the outflow than is seen in
the starburst disk. The average spectrum of the wind has smaller 6.2/7.7 ratios and larger 11.3/7.7 ratios than the average spectrum of the starburst,
suggesting the grains may be larger and more ionized, on average, than
those in the starburst disk. 

\subsection{H$_2/PAH$ variations in the M82 wind}

\begin{figure}
\includegraphics[width=0.45\textwidth]{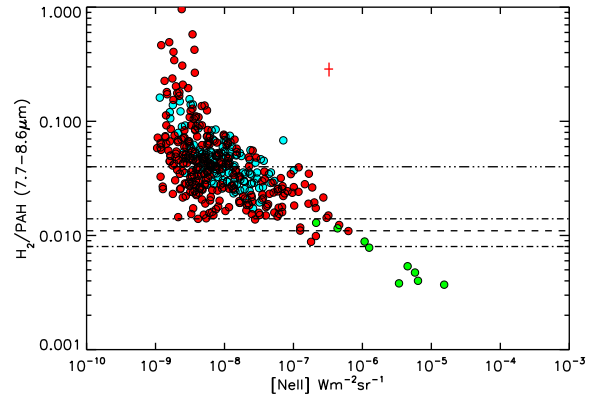}
\caption{Plot of the H$_2$ S(1)-S(3)/PAH ($7.7+8.6\mu$m) as a function of the [NeII] surface brightness. Each datapoint represents a $10\arcsec\times10\arcsec$ region. Red points are the northern outflow, cyan the southern outflow, and green the central 1 kpc of M82 from \citet{beirao08}. The dashed line and the dash-dotted lines represent the average and the $1\sigma$ variation, respectively, of the H$_2$ S(1)-S(3)/PAH ($7.7+8.6\mu$m) ratio in the star-forming galaxy sample of \citet{roussel07}. The dotted-dashed line represents the upper limit in the H$_2$ S(1)-S(3)/PAH ($7.7+8.6\mu$m) ratio predicted by PDR models \citep{guillard12}.} The error bar in red represent typical uncertainties in the northern outflow, while the typical uncertainties in the data for the southern outflow region are smaller than the points.
\label{h2pahfig1}
\end{figure}

In Figure~\ref{h2pahfig1} we plot the variation of H$_2 S(1)-S(3)$/PAH$7.7+8.6\mu$m
as a function of [NeII] surface brightness. Since the [NeII] emission arises predominantly from the starburst disk, this traces the relative H$_2$ and PAH emission as a function of distance from the starburst. This ratio has been previously used by \citet{ogle10} and \citet{guillard12} to show the relative importance of the UV versus mechanical heating of the gas. 
We can see that the H$_2/PAH$ ratio decreases with increasing [NeII] flux. The vast majority of the mapped regions lie well above the values seen in star-forming galaxies in the local Universe (the dashed lines - see \citet{roussel07}). The H$_2/PAH$ ratio values span two orders-of-magnitude, 0.01-1. Most of the M82 wind points have H$_2/PAH$ between 0.01 and 0.10, but $\sim4$\% have extreme values, H$_2/PAH$ $> 0.5$. The fact that the H$_2/PAH$ ratios seen in the wind are well above those seen in the SB disk, the normal galaxies observed by \citet{roussel07}, and PDR models of \citet{guillard12} (see section 4) suggests that the H$_2$ is not solely excited by stellar radiation from the starburst.

\begin{figure*}
\includegraphics[width=0.45\textwidth]{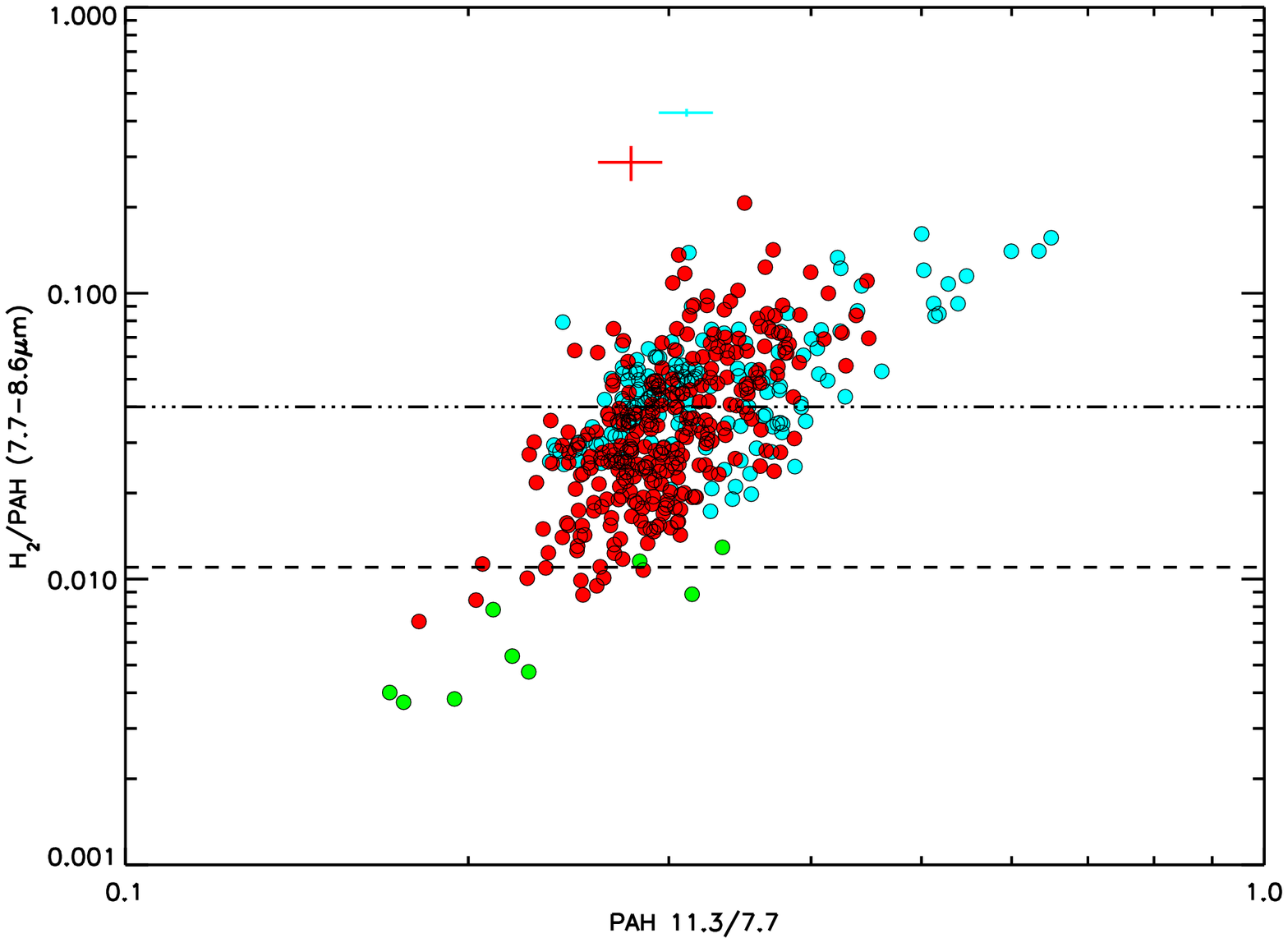}
\includegraphics[width=0.45\textwidth]{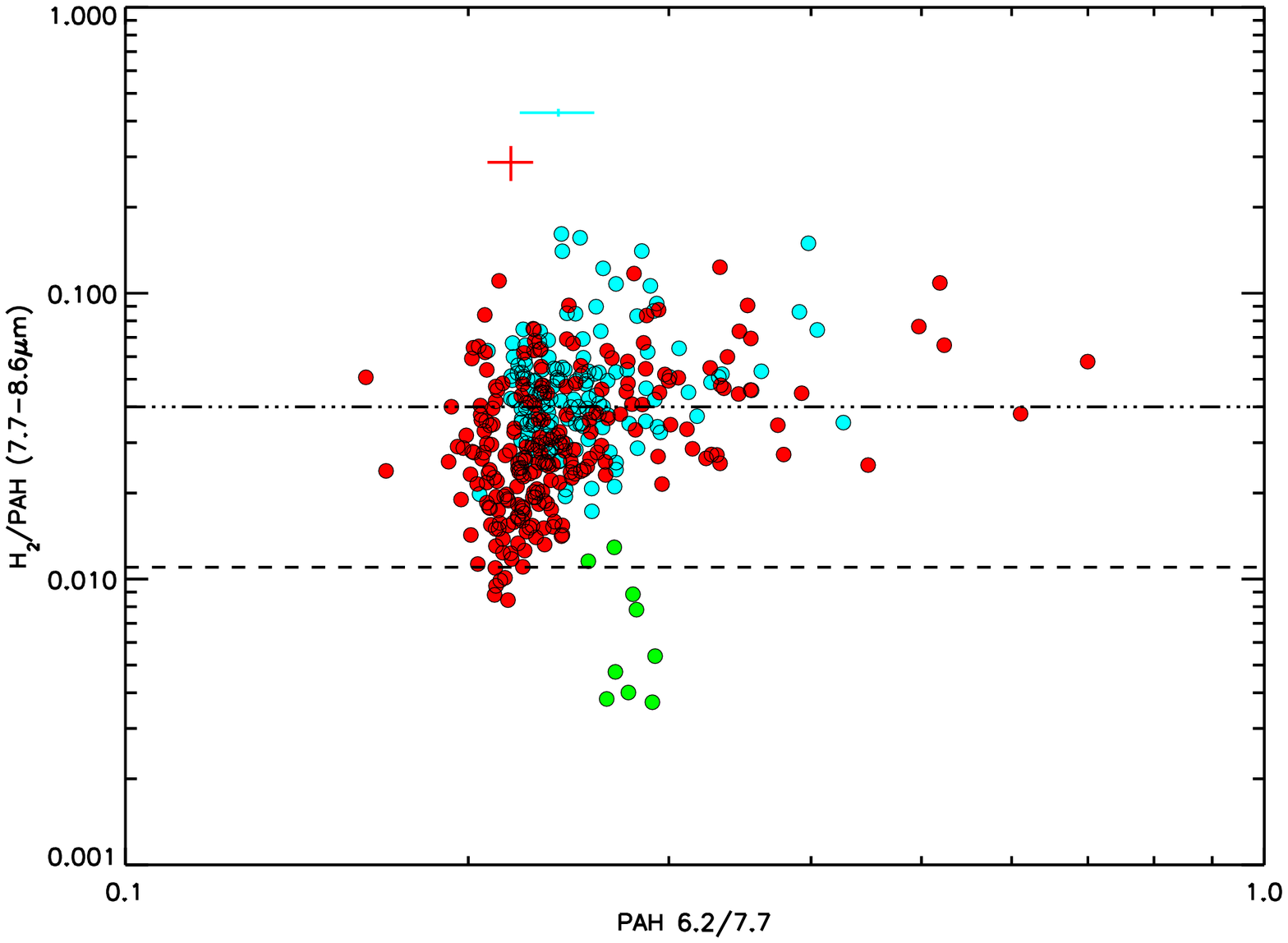}
\includegraphics[width=0.45\textwidth]{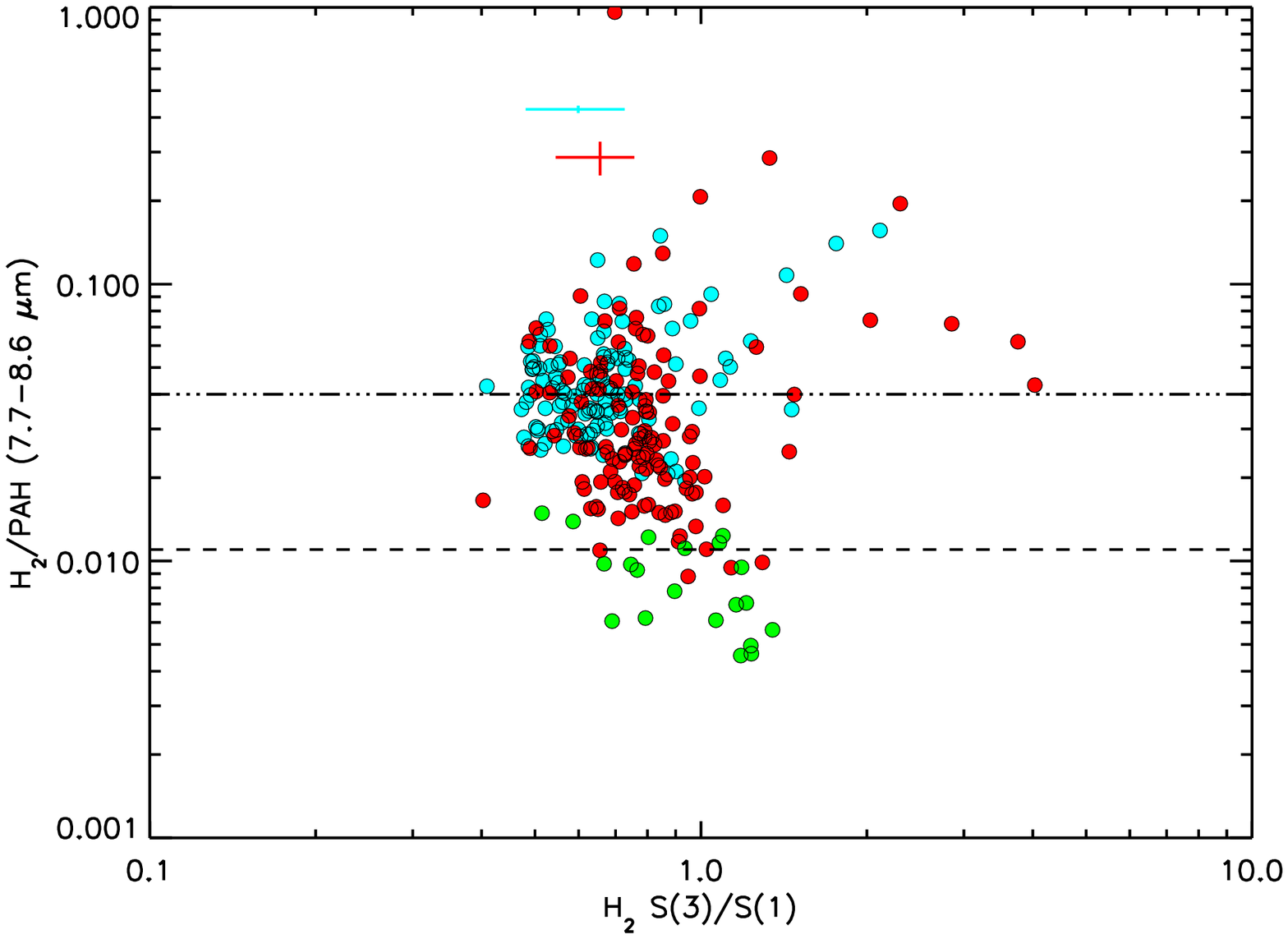}
\includegraphics[width=0.45\textwidth]{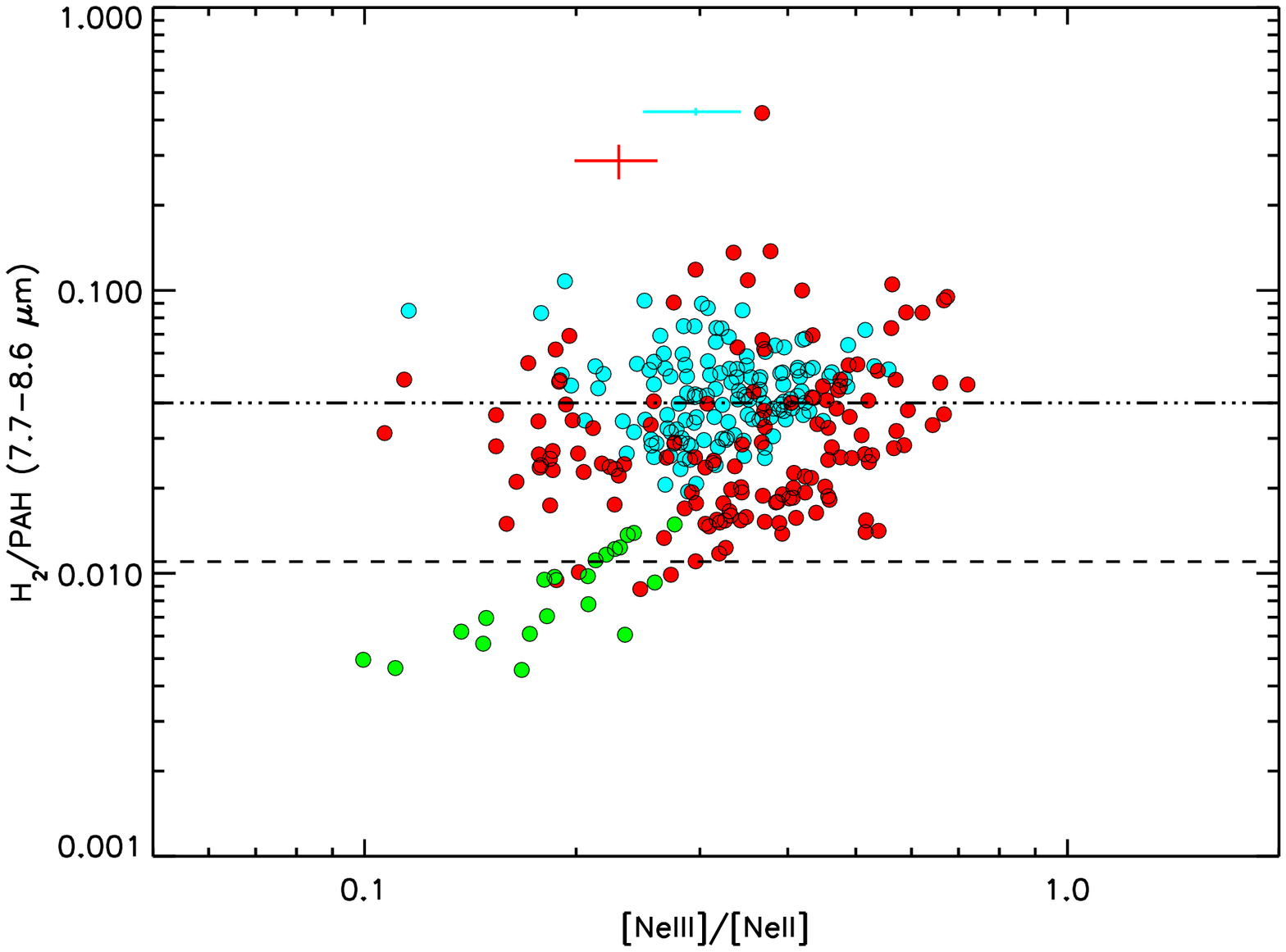}
\caption{Plot of the H$_2$ S(3)-S(1)/PAH ($7.7+8.6\mu$m) as a function of PAH 11.3/7.7 (upper left), PAH 6.2/7.7 (upper right), H$_2$ S(3)/S(1) (lower left) and [NeIII]/[NeII] (lower right). Each datapoint represents a region of $10\arcsec\times10\arcsec$ in extent. Red points are the northern outflow, cyan the southern outflow, and green the central 1 kpc of M82 from \citet{beirao08}. The error bars represent typical uncertainties in the northern (red) and southern (cyan) outflows. The dashed lines represent the average and the $1\sigma$ variation of the H$_2$ S(1)-S(3)/PAH ($7.7+8.6\mu$m) ratio in the star-forming galaxy sample of \citet{roussel07}. The dotted-dashed line represents the upper limit in the H$_2$ S(1)-S(3)/PAH ($7.7+8.6\mu$m) ratio predicted by PDR models \citep{guillard12}.}
\label{h2pahfig2}
\end{figure*}

We can also investigate possible correlations of the H$_2/PAH$ ratio with diagnostics of the atomic gas and the size and ionization state of the grains. 
In Figure~\ref{h2pahfig2}, we show H$_2/PAH$ as a function
of the PAH 11.3/7.7 ratio, the PAH 6.2/7.7 ratio, the H$_2$ S(3)/S(1) ratio, and the [NeIII]/[NeII] ratio. We see that in
the disk the H$_2/PAH$ ratio increases with increasing PAH 11.3/7.7
ratio. While the trend of increasing H$_2/PAH$ in the wind is similar to that seen in the disk, the wind values extend to much larger values of both 11.3/7.7 PAH and H$_2/PAH$ than seen in the starburst. The scatter to high H$_2/PAH$ and a much larger range in the 6.2/7.7 ratio is seen in the upper right panel of Figure~\ref{h2pahfig2} as well. In the lower left, we observe an anti-correlation between H$_2/PAH$ and S(3)/S(1) in the disk as warmer gas has lower H$_2/PAH$. The wind points seem consistent with this trend but the uncertainties are large. This general lack of very warm molecular gas and high H$_2/PAH$ ratios in the wind may be a function of inability of the MIR diagnostic features to probe the hottest gas.
The [NeIII]/[NeII] also increases with H$_2/PAH$ in the disk, and the wind points to the North seem consistent with the trend, but clearly there are many wind points in a few regions in the northern outflow which have much higher H$_2/PAH$ ratios than seen in the disk at a given [NeIII]/[NeII] value.

\begin{figure*}
\includegraphics[width=0.45\textwidth]{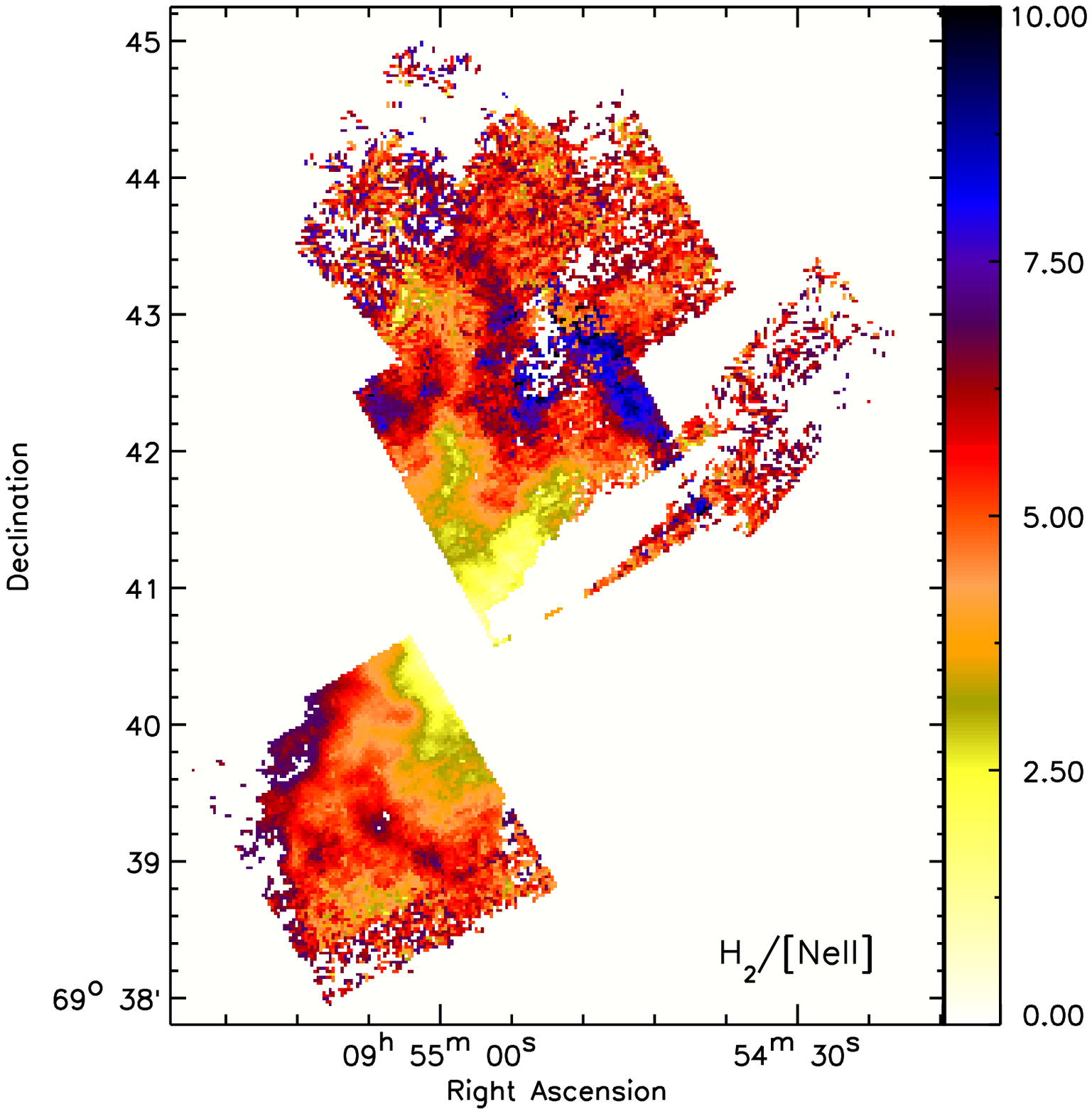}
\includegraphics[width=0.45\textwidth]{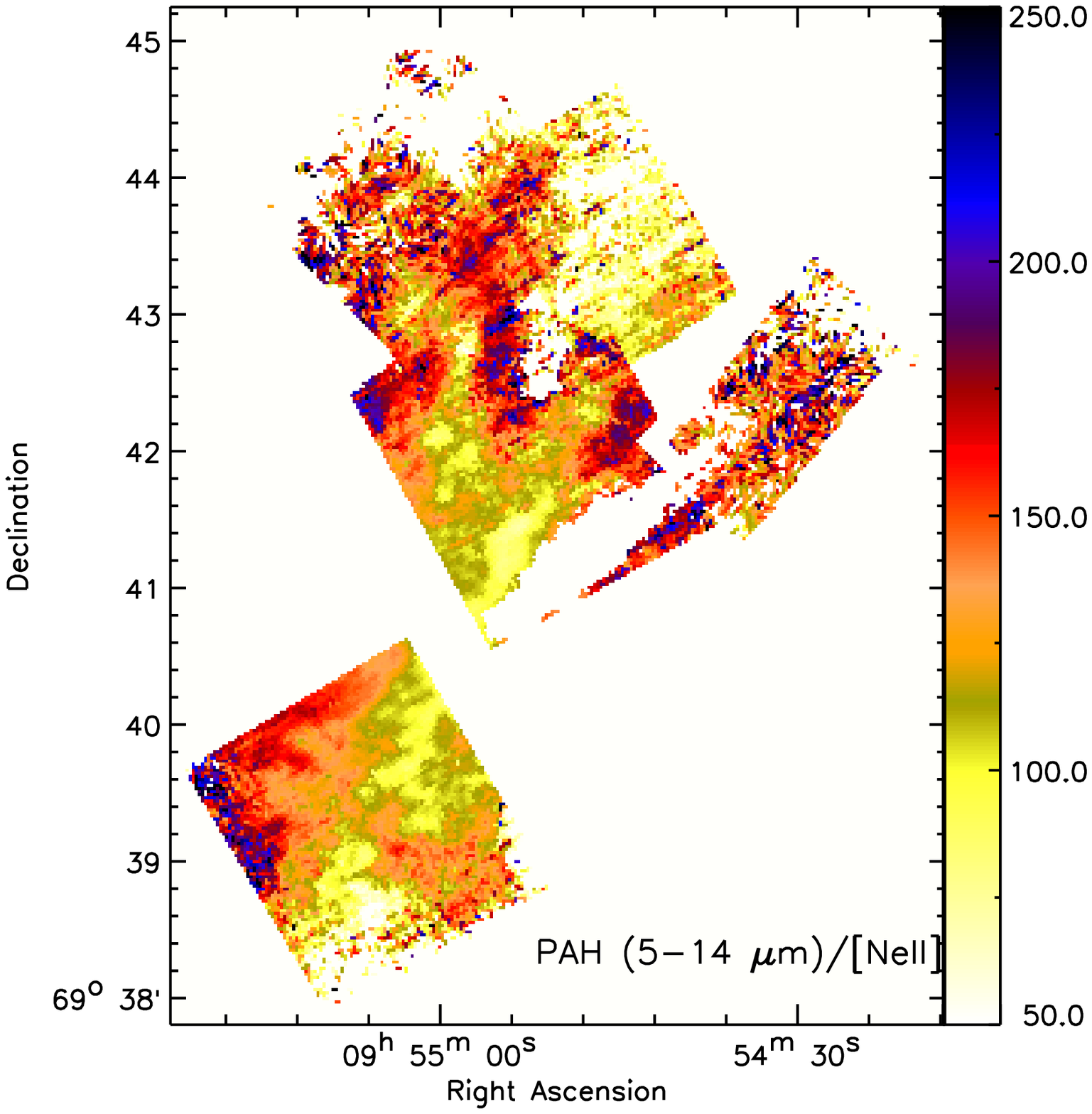}
\caption{Maps of the H$_2 S(1)-S(3)$/[NeII] ratio (left),
PAH$5-17\mu$m/[NeII] (right). H$_2$ is the sum of the S(1), S(2), and S(3) H$_2$ line fluxes and PAH$5-17\mu$m is the sum of the fluxes of all PAH features between $\sim5$ and 17$\mu$m. All maps have been clipped at the $3-\sigma$
surface brightness level. The white radial strip in the bottom two panels
are regions that have been masked out due to scattered light in the LL2
module, as in Figures~\ref{pahmapfig} and~\ref{ratiomapsfig}.}
\label{neratiomapsfig}
\end{figure*}

To further help us understand what drives the variations of the
H$_2/PAH$ flux ratios, we present maps of H$_2$/[NeII] and
PAH/[NeII] in Figure~\ref{neratiomapsfig}. In general, the regions with high H$_2/PAH$ correspond to high PAH/[NeII], suggesting that it is not simply a decreasing PAH that causes the rise in H$_2/PAH$ off the plane of M82.  Only in the northern edge of the map do we see a lower PAH/[NeII] ratio, which might indicate reduced PAH heating efficiency. It is important to note, however, that while the PAH/[NeII] ratio does change by factors of a few across the map, this is much less than the variation observed in H$_2/PAH$. Therefore the latter cannot be solely explained by a change in the strength of the PAH emission alone.

\subsection{H$_2$ masses and temperatures}
\label{h2sec}

\begin{figure*}
\includegraphics[width=0.43\textwidth]{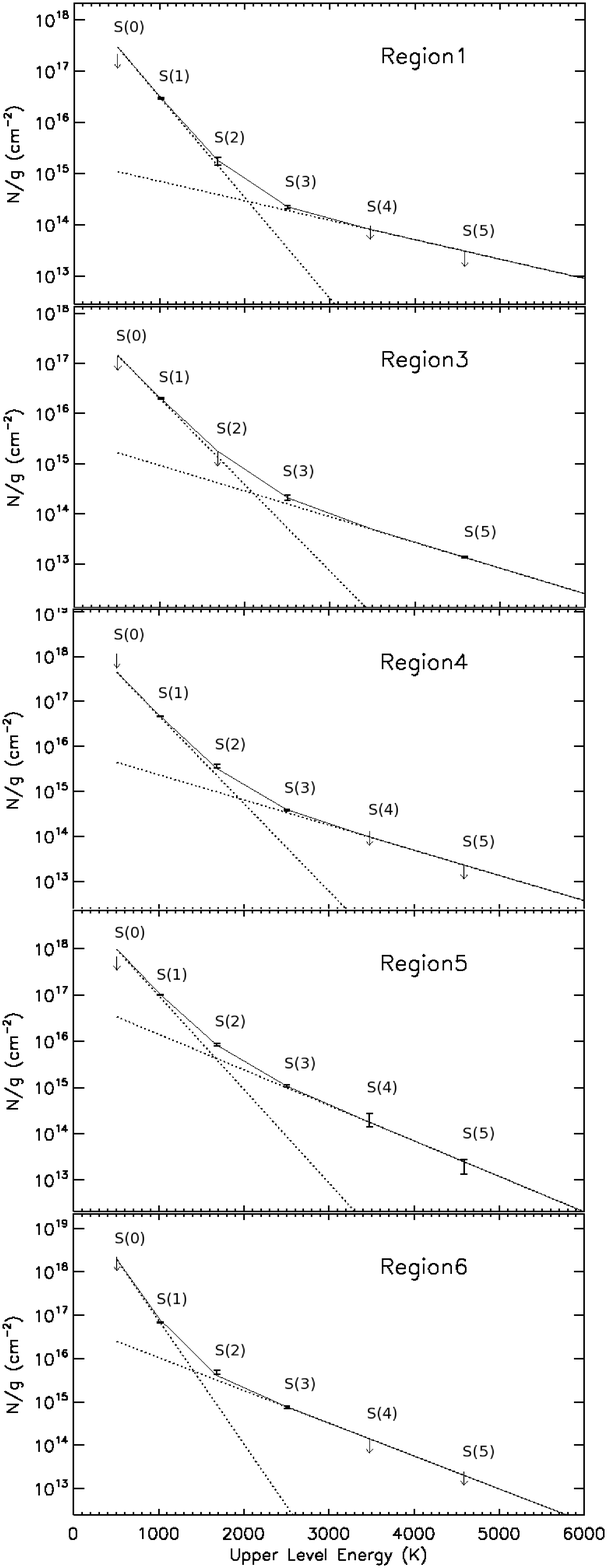}
\includegraphics[width=0.43\textwidth]{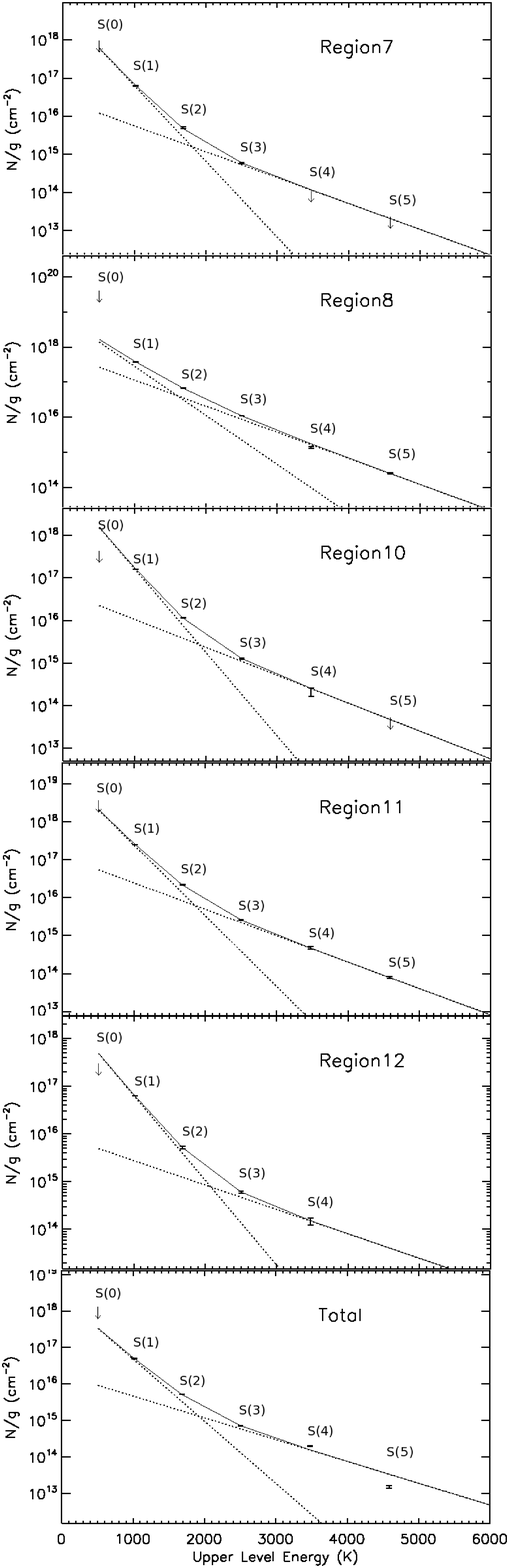}
\caption{H$_2$ excitation diagrams of most selected regions. The dotted lines show a 2-temperature LTE fit of the data. The upper limits of the H$_2$ lines represent a 3-$\sigma$ noise level at the corresponding line.}
\label{h2fig1}
\end{figure*}

With the rotational H$_2$ emission line fluxes in Table~\ref{h2fluxtab}, we can
estimate the gas temperature, column density, and mass distributions,
assuming that the ortho/para ratio is in local thermal equilibrium (LTE).
In Figure \ref{h2fig1} we plot the excitation diagrams for all regions,
except 2 and 9, which did not have sufficient signal-to-noise for fitting
the data with these sorts of models. Using the Boltzmann equation, we fit the H$_2$
lines with two components, each with different temperatures. We then derive the total column density, H$_2$ temperatures
and mass distributions. 
In fitting the data we set the minimum allowed temperature to 100 K and treated every upper limit as a 3-sigma detection. The S(1) line is the line with the highest signal-to-noise. Therefore, when we remove the S(0) line as a detection, we obtain essentially the same results.

The diagrams for Regions 1, 3, 4, 5, 7 and 11 look very similar in shape, derived
from very similar temperatures for each component. In Region 8, both
components have similar temperatures, while in Region 10, there is a large difference in temperature between the two components. We note that Regions 8 and 10 have the most discrepant S(0) upper limits relative to the other fits, making the mass estimates particularly uncertain.
In Table~\ref{h2tab1} we present the properties of the warm H$_2$ gas
in these regions of interest. In all the regions the temperature of
the warmer component is above 500 K, especially warm when compared with
typical star forming regions \citep[e.g.][]{roussel07}. The column density of the bulk of the warm H$_2$ gas ranges by a factor of $\sim30$. The mass ratio
between the warmer and colder component varies between $0.0003-0.02$, with the lowest
value observed in Region 8 and the highest in Region 1. The region with
the most luminous H$_2$ emission is Region 8, where the warmer component
accounts for $\sim95\%$ of the luminosity, whereas the next most luminous region, 11,
the colder component accounts for $\sim65\%$ of the luminosity. 

We also integrated the emission over the entire observed nebula surrounding M82 
and measured the line fluxes, which are included in Table~\ref{h2fluxtab}. The total warm H$_2$ mass derived from the integrated emission is $\sim4.8\times10^6$ M$_{\odot}$.
For completeness, we note that \citet{veilleux09} finds a mass of
H$_2$ at a temperature T$>$1000 K (thus hotter than our ``hot'' gas) of 1.2$\times$10$^4$ M$_{\odot}$,
based on near-infrared observations. The relatively low mass is
not surprising since the near-infrared ro-vibrational lines are probing
very high excitation gas which emits very efficiently in these lines. The mass of warm molecular gas in the wind that we are detecting with the IRS is thus more than a factor of 100 more than is estimated from the NIR line maps. We have also estimated the H$_2$ masses from fitting the H$_2$ line fluxes with molecular shock models (see Section 4.3).

Considering a cold molecular gas mass of $\sim10^9$M$_{\odot}$ \citep{walter02,salak13}, we calculate a ratio of warm to cold molecular gas in the outflow of about 0.01.
Therefore the warm molecular gas we measure here is a relatively minor component of the total molecular mass in the ISM. Care must be taken when making this comparison
as the emission regions for the CO and H$_2$ do not overlap completely.
In fact, the most extended CO emission is only within a few arc minutes
of the plane and much less extended \citep{salak13,walter02}, although new studies trace the CO-emitting gas to larger distances \citep{leroy14}. However,
we note that most of the estimated mass for the warm molecular gas in
our observations also comes from regions which lie close to the disk
(and likely the base of the wind).

\begin{figure}
\includegraphics[width=0.45\textwidth]{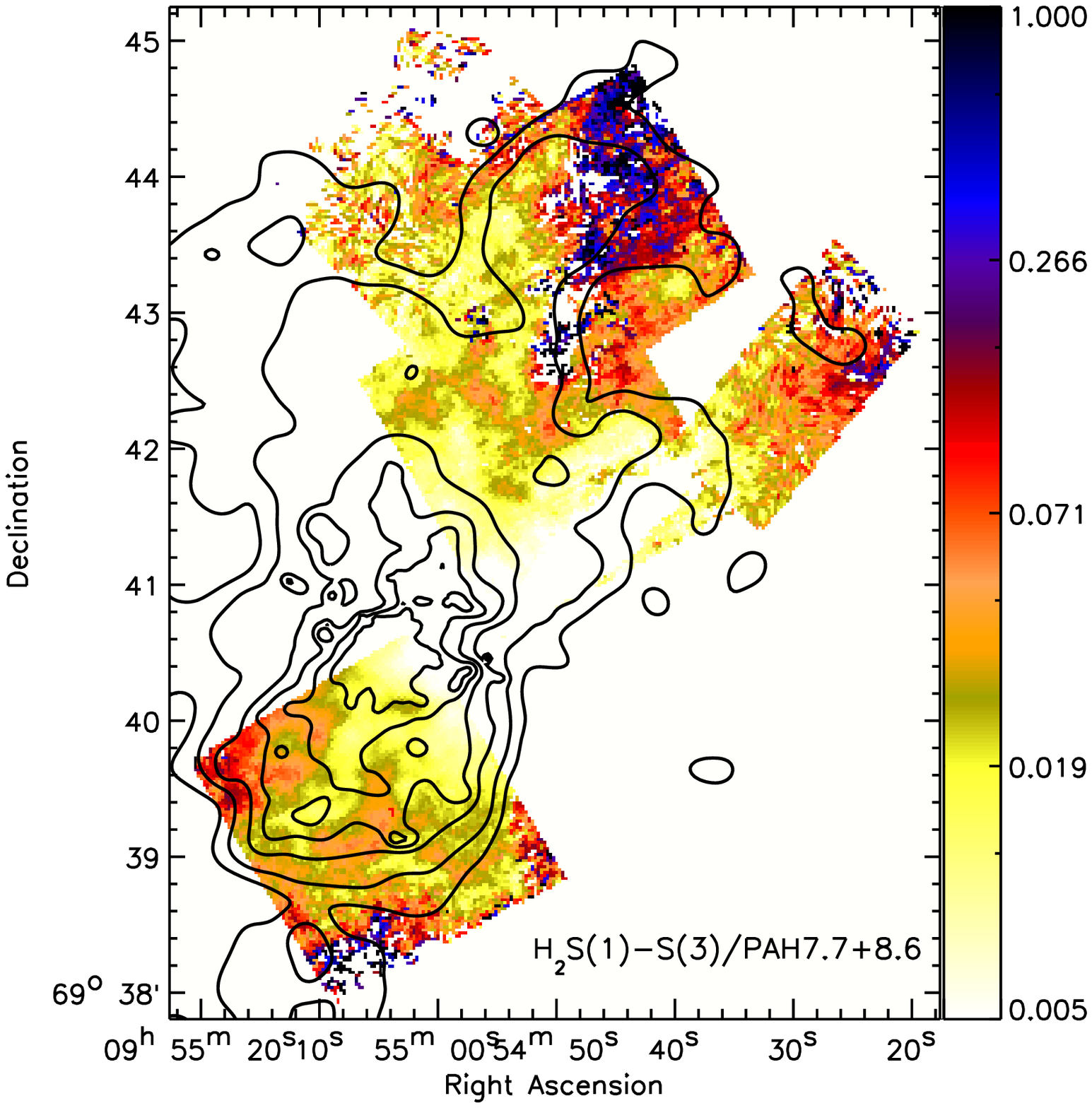}
\includegraphics[width=0.45\textwidth]{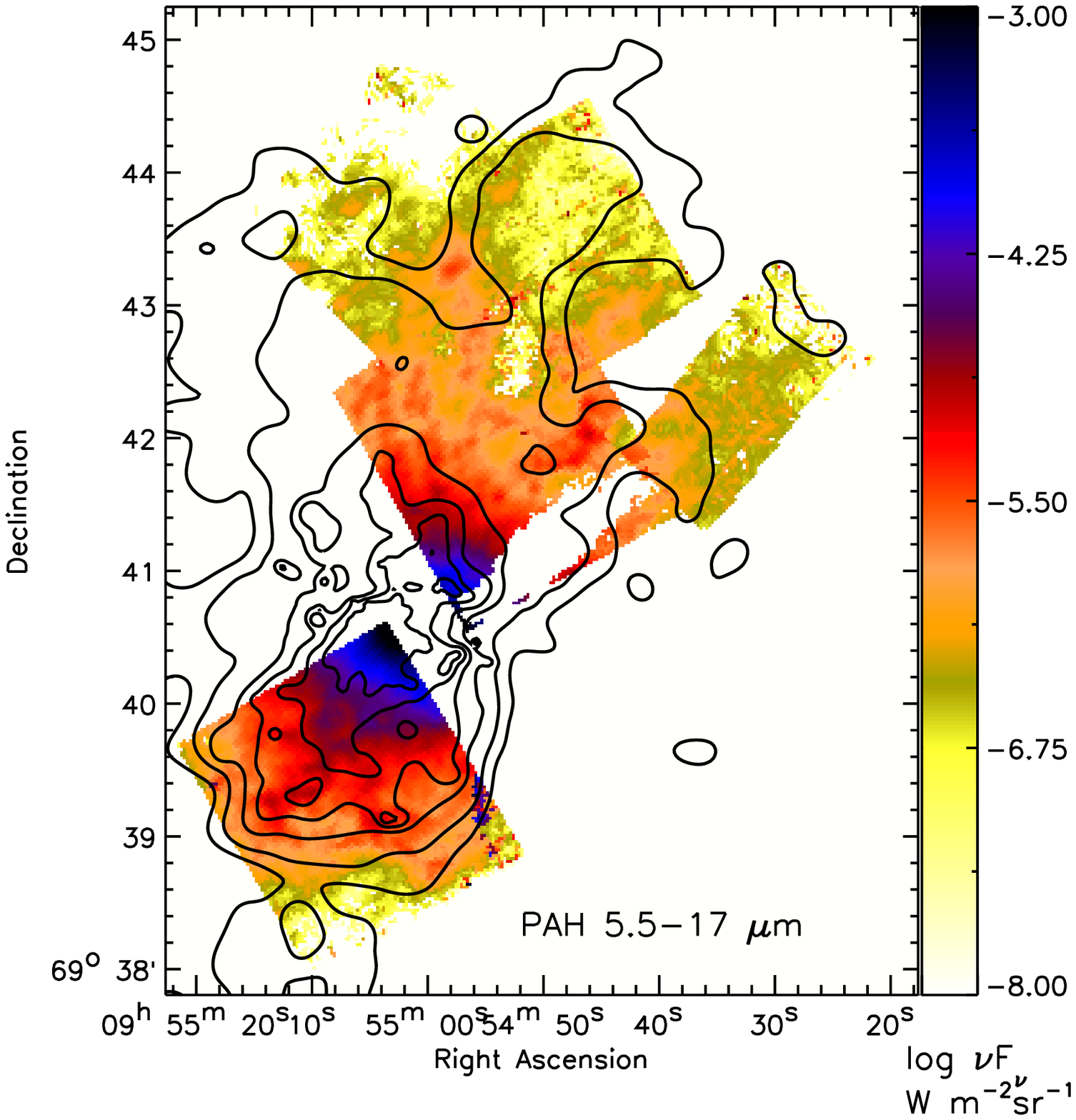}
\caption{Image of the H$_2/PAH$ (top) and total PAH flux
($5.2-17\mu$m) (bottom), overlaid with contours from soft ($0.3-1.6$
keV) X-ray emission map of M82 (Strickland \& Heckman 2009).}
\label{xrayfig}
\end{figure}

\section{Discussion}
\label{discussion}

\subsection{H$_2$ heating processes}

In the scenario of thermal excitation of H$_2$, three mechanisms dominate the heating: (1) UV radiation from starburst, (2) X-rays
from the starburst and wind, and/or (3) shocks induced by the outflow.
Following \citet{ogle10}, we use the H$_2$-to-PAH luminosity ratio
to investigate the contribution of UV photons to the total heating of
the H$_2$ gas in the outflow. In starburst galaxies like M82, we expect the H$_2$ emission to be predominantly powered
by UV radiation from young stars but undoubtedly, processes such as
low-velocity shocks and turbulent dissipation contribute to the H$_2$ emission. 
It should also be noted that H$_2/PAH$ ratios on the order of 0.01 have also been seen from translucent clouds in the cold, diffuse ISM of the Milky Way, which are thought to be excited by UV photons \citep{ingalls11}.

There is a clear offset of the wind points from those in the
SB disk in terms of H$_2$, PAH and ionized atomic emission (Figs.~7, 8, \&
9). The data in regions of the outflowing gas show a large range in both
[NeIII]/[NeII] and H$_2/PAH$, but are clearly offset to higher H$_2/PAH$
at a given [NeIII]/[NeII] ratio compared to regions within the SB disk,
suggesting an additional excitation mechanism at work. In fact, models suggest that the large values of the H$_2/PAH$ ratios
observed in some regions ($>$0.1) cannot be explained by
photo-ionization \citep{guillard12}.  

The H$_2$ gas could be indirectly excited by X-rays emitted from
the expanding bubble of hot and tenuous gas generated by the intense
star formation within M82 \citep{begel89}, high velocity shocks in the
entrained gas, X-rays due to inverse Compton scattering of IR photons.
In M82, the X-ray emission
covers approximately the same region as the optical emission line filaments
\citep{watson84, shopbell98} and radio emission \citep{seaquist91}.
For example, the high H$_2/PAH$ in the northern tip $\sim$5 kpc from the
disk, appears to be associated with a region with X-ray emission (Figure~\ref{xrayfig}). 

We can use the total H$_2 S(1)-S(3)$ luminosity and the total X-ray
luminosity in the outflow to calculate the H$_2$-to-X-ray luminosity
ratio \citep{stevens03}.  The H to H$_2$ transition can occur at
relatively low column densities, N$_{H}\approx$10$^{21}$ cm$^{-2}$.
At such low columns, even relatively soft X-rays can penetrate.  Our
estimates for the M82 outflow are $L(H_2 S(0)-S(3))/L_X (2-10$ keV$)\sim1$
\citep{stevens03} and $L(H_2 S(0)-S(3))/L_X (0.2-4 $keV$)\sim0.5$ \citep{watson84,strick07}, over the northern and southern $IRS$ map. 
Using XDR models \citep[e.g.][]{maloney96} we can estimate whether the H$_2$ line emission
could be powered by X-ray heating, following the procedure described in \citet{ogle10}. In these models, $30\% - 40\%$ of the
absorbed X-ray flux goes into gas heating via photoelectrons, and the cooling by H$_2$ rotational lines in XDR models is $\sim2$\% of the total gas cooling for a gas temperature of 200 K.
At this temperature, the ratio of the first four rotational lines to the total rotational line luminosity is L(H$_2$ 0--0 S(0)--S(3))/ L(H$_2$) = 0.6. Combining the above factors, we estimate a maximum H$_2$ to X-ray luminosity ratio of L (H$_2$ 0--0 S(0)--S(3))/ L$_X$ (2--10 keV) $< 0.01$ \citep{guillard12b}. 
This upper limit on the H$_2$-to-X-ray luminosity is conservative because it assumes
that  all the X-ray flux is absorbed by the molecular gas, which is
likely not the case (since the medium in which X-rays propagate is inhomogeneous).  Therefore, we conclude that X-rays are not the dominant source of heating
of the H$_2$ gas in the M82 wind, since the observed ratios are well above those found in XDR models. This is also seen in some extreme environments, such as powerful radio galaxies, where H$_2/PAH$ $\sim 0.5-1$ and the H$_2$/X-ray ratio also exceeds that expected from XDRs alone \citep{ogle10}.

Ambient gas clouds heated by the mechanical energy of the wind and dense shell
fragments from the wind bubble carried along by the wind are the two main
sources of X-ray emission in the wind \citep{strick97}.  In Figure~\ref{xrayfig}
we show the contours of soft X-ray emission from the outflow of M82
overlaid on the H$_2/PAH$ and total ($5-17\mu$m) PAH maps. The X-ray
and PAH maps show a remarkable correlation, particularly in the
southern part of the bi-conical outflow, although both maps are projections of the true 3D structure and the precise locations of the hot gas and dust grains are unknown. 
Moreover, there are regions of
high surface brightness X-ray emission with relatively little PAH emission
(in the north along the outer wind axis).
We also see that the middle
lobe of X-ray emission in the northern outflow, at about 5 kpc from
the disk, appears related to the increase of H$_2/$PAH$7.7+8.6\mu$m surface brightness.
The morphology of H$_2/PAH$ closely resembles the
morphology of the H$_2$ ($2.12\mu$m)/PAH maps in \citet{veilleux09}. The
regions where H$_2/PAH>0.1$ correspond to regions
where H$_2$ ($2.12\mu$m)/PAH$>0.003$. 

While there is a reasonable spatial correlation between the
surface brightness distribution of the soft X-ray and H$_2$ emission, 
a correlation is apparent between soft X-rays
surface brightness and H$_2/PAH$ ratio distributions only in high surface brightness X-ray emitting regions. 
There are several
possibilities why this might be so, but the most obvious is that both the
X-ray and H$_2$ emission are fundamentally powered by the same mechanism,
namely, the very hot plasma that is acting as the ``piston'' driving the
wind and heating clouds entrained in the wind. However,
the warm H$_2$ and hot X-ray emitting plasma should not be co-spatial on
small scales, and instead the H$_2$ is likely seen in projection against the hot
wind, along the edges of the cone in dense knots of gas, heated by shocks
(see below). Such a situation would also arise in gas with a wide range of
densities or through changes as individual clouds are shredded as they are
accelerated by the action of the piston \citep[][]{cooper08}. This is also
shown by the strong correlation between the X-rays and the PAH emission
(Figure~\ref{xrayfig}b), especially in the Southern wind. As the PAHs and H$_2$ molecules 
cannot be expected to survive within the $10^6-10^7 K$ X-ray emitting plasma, it is most likely that they lie along the edges of this phase of the outflow. So
while the power source is likely the same, the two emission regions,
not surprisingly, represent different physical regions and gas states
being impacted by the driving very hot plasma \citep{strick07}.

\begin{figure*}
\includegraphics[width=0.45\textwidth]{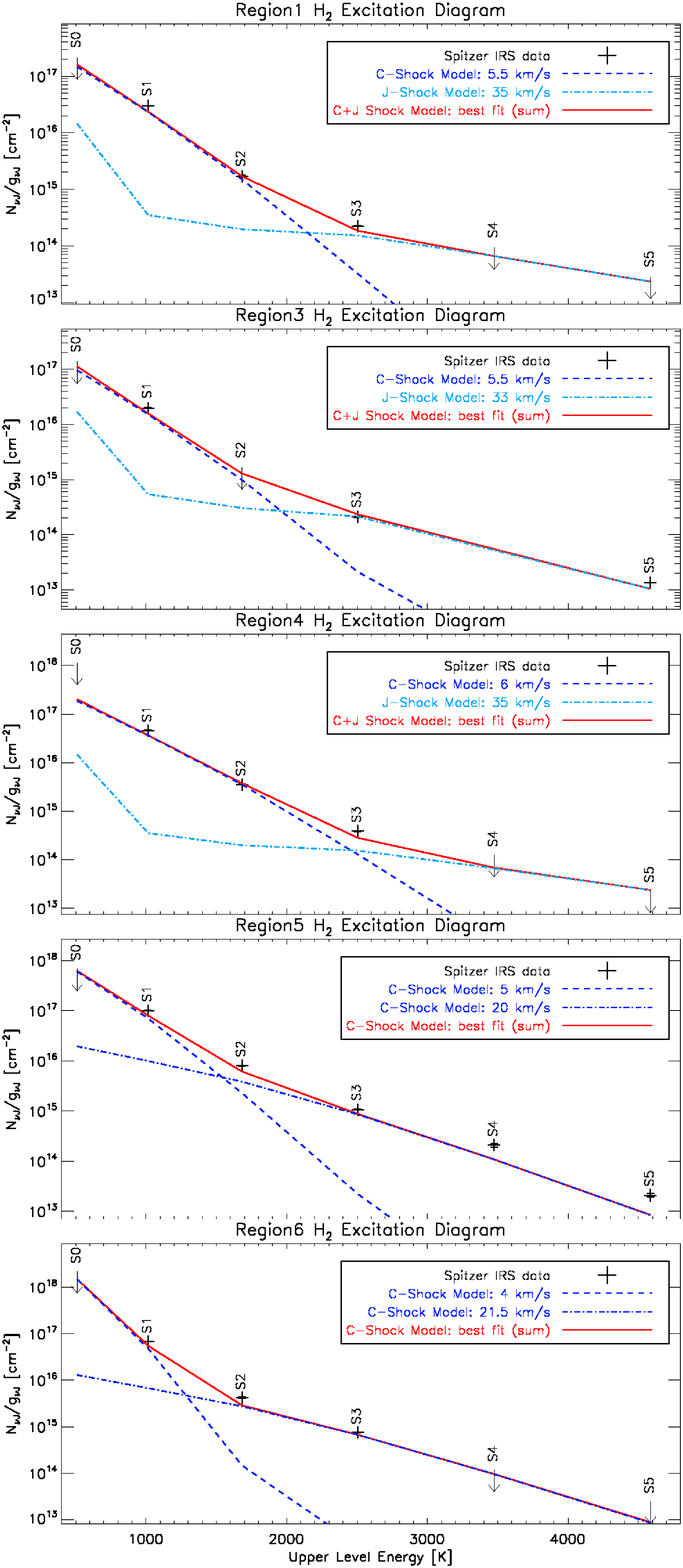}
\includegraphics[width=0.45\textwidth]{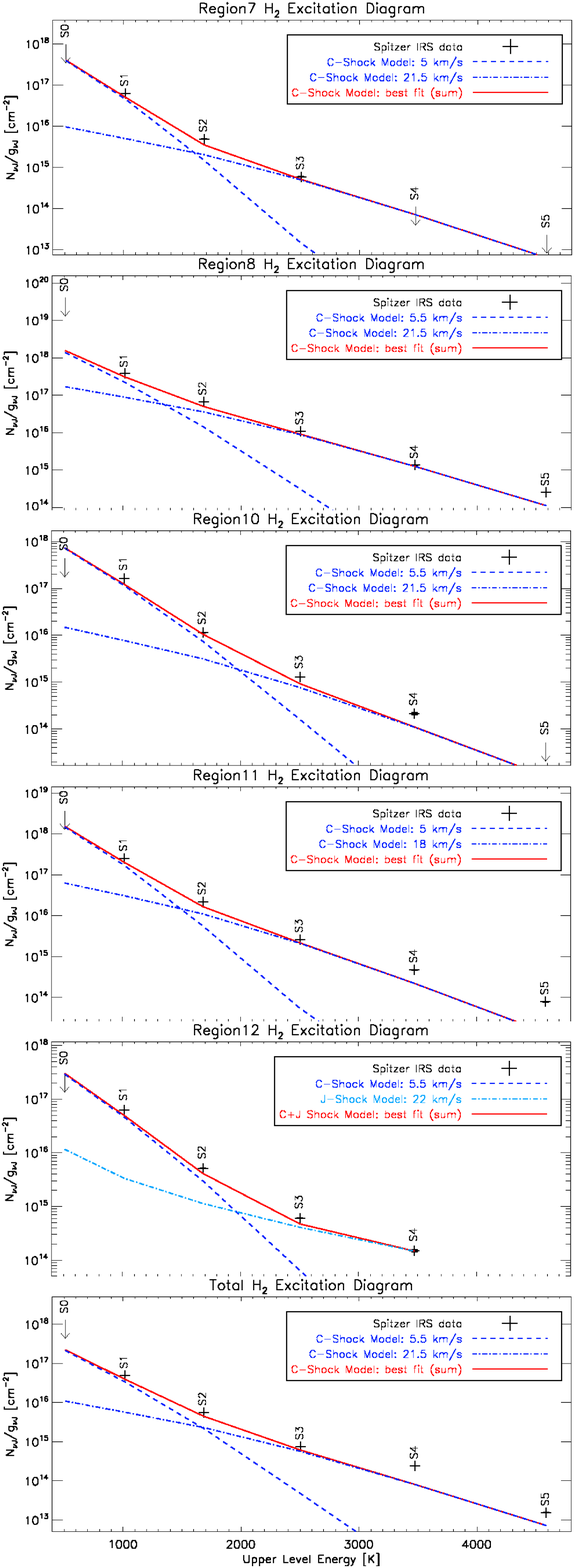}
\caption{Shock excitation diagrams of most selected regions in the outflow.
The best fit models for either J- (light blue) or C-shocks (dark blue) or a combination of each are
indicated in the legend to each panel (see text for details).  The sum of
or the individual the best fit models is shown as a solid red line.
The level populations of the various rotational levels of the (0-0)
H$_2$ lines are indicated by plus signs if they are detection or downward
pointing arrows for upper limits and are labeled as S(1), S(2), S(3), S(4), and S(5).}
\label{h2figshock}
\end{figure*}

\subsection{The velocities and kinetic energy of the molecular gas}

Kinematics of the gas can be used to determine if the different gas phases
observed in the outflowing wind are related.  If the observed kinematics
of the warm molecular gas as probed by the H$_2$ rotational lines is
similar to that of the cold molecular gas as probed by the CO lines, the
IR atomic neutral and ionized emission lines, or the lines of the warm
ionized medium, then this might suggest a direct physical relationship
between these phases. 

\citet{salak13} derived a velocity of 160 km/s for the CO(1-0) emitting gas for the northern
outflow region. However, the farthest extent of the CO emission is
only about 1 arc minute north of the disk of M82, while the IRS map extends to 4 arcmin. The
velocity derived by \citet{salak13} is not very different from the velocities in the
neutral and ionized atomic gas in the infrared within the inner wind region
$<1$ arcmin from the disk \citep{contursi13}. 
If we restrict ourselves to the inner wind region within a few arc minutes of
the plane of M82, the velocities in the warm ionized gas are only a few 100 km s$^{-1}$ or less
within a couple of arc minutes of the disk of M82 \citep{shopbell98}.
These velocities are consistent with those
observed in IR neutral and ionized atomic lines \citep{contursi13}. However, farther out of the
plane of M82, much higher outflow velocities ($\sim$600-800 km s$^{-1}$)
are observed in H$\alpha$ for the outflow in M82 \citep{shopbell98}. Line splitting
is observed in both the CO and H$\alpha$ with similar velocities \citep{walter02}, making the association between the cold
molecular gas and the warm ionized gas stronger \citep[cf.][]{salak13,shopbell98}.  

Such a direct relationship would be consistent
with a picture where the emission we observe in the extended outflow of
M82 is mainly from entrained and swept-up clouds that are predominately
shock-heated by a high ram-pressure, hot plasma generated within
the starburst region. If we envision the regions of warm H$_2$ emission
as being the areas in the clouds that are shock-heated by the expanding
plasma, we would expect these phases to co-exist with the warm ionized
gas \citep{heckman90} and the
cold/warm molecular gas.

Assuming that the velocity of the warm H$_2$ is similar to the deprojected velocity of the cold molecular gas derived by \citet{salak13}, we can calculate its total kinetic
energy as $E_K=0.5\times M_{H_2}v_{outflow}^2\sim5.9\times10^{53}$ erg for
$M_{H_2}\sim4.8\times10^6 M_{\odot}$ and $v_{outflow}$=160 km s$^{-1}$.
Our estimates are considerably higher
than those of \citet{veilleux09} because both their mass estimates are
a factor of 500 less and they assumed a velocity of 100 km s$^{-1}$
for the warm molecular emission. However, they are smaller compared to
those estimated for the cold molecular and warm ionized medium. \citet{salak13}
estimate that the kinetic energy of the wind in the cold molecular gas
is $\approx 1-4\times$10$^{56}$ erg. \citet{walter02} estimate a kinetic energy of $\approx$3$\times$10$^{55}$ erg, but they do not map the
CO distribution out to large distances from the disk compared to \citet{salak13}.

\subsection{H$_2$ supersonic turbulent heating and shock models}

The hot
plasma driving the outflow will likely transfer part of its ram
and thermal pressure to swept-up and entrained dense molecular
gas by driving shocks and by turbulent mixing between hot and cold
gas  \citep[see][]{guillard09}.  To model such shocks, we use a grid
of stationary shock models computed with an updated version of the
\citet{lesaffre13} code to derive the physical shock parameters that
explain the H$_2$ line ratios.  Our initial conditions are that the gas is
diffuse, magnetized with a pre-shock magnetic field of intensity $B[\mu
G] = \sqrt{n_H[\rm cm^{-3}]}$, and irradiated with a radiation field
of $G_0 = 1$ in Habing units. As the ``Habing field" would photodissociate the H$_2$, we assume some self-shielding. In the models, we integrate the extinction along the model from the pre-shock, and we take into account the shielding and self-shielding, which is parametrized in the photo-dissociation reaction rates of H$_2$ and CO \citep{lesaffre13}. At a density of $10^4$ cm$^{-3}$, the ionization fraction in the pre-shock gas is $6\times10^{-8}$. This matches
approximately our estimate of the ionization fraction of $10^{-7}$ at 10 kpc
from the X-ray source. We fit the H$_2$ line fluxes observed in the
different regions with one shock or a combination of two shock velocities
depending on how many H$_2$ lines are detected, for three different
pre-shock densities: $n_H = 10^2$, $10^3$ and $10^4$~cm$^{-3}$, (see
\citet{guillard09} for details of the method). The results of these fits
are shown in Figure~\ref{h2figshock}, where we present the H$_2$ excitation
diagrams for the best velocity combinations (lowest $ \chi ^2 $), and
for the intermediate pre-shock density of n$_{\rm H}$ = 10$^3$ cm$^{-3}$. We present in Table~\ref{h2tab2} the results of the shock models for each region and the integrated spectrum (labeled ``Total''). 
The total gas mass derived from the shock models is $M_{H_2}=1.7\times10^7 M_{\odot}$, which results in an estimate of the kinetic energy of $E_K\sim2\times10^{54}$ erg. This is a factor of 3 higher than the estimates calculated in the temperature diagrams of Figure~\ref{h2fig1}. This difference arises due to the fact that we are integrating the postshock temperature profile down to 50 K (where the H$_2$ mid-IR lines start to emit), and because local thermodynamic equilibrium (LTE) is not assumed in the shock models.

For all regions, the best-fit shock velocities are in the range 4-35
km s$^{-1}$ (grid spans 3-40 km s$^{-1}$; Tabl~\ref{h2tab2}). These
velocities are much lower than the estimated bulk outflow velocity
\citep{lehnert99}. 
The regions show a range of best fitting models, but typically the best fitting
pre-shock densities are n$_{\rm H}$=100-1000 cm$^{-3}$. Our modeling
is not intended to be exhaustive, but to simply provide an estimate of the physical parameters in
the warm molecular gas of the outflow.

At the range of preshock densities and magnetic field strengths
considered, the C-shocks turn into J-shocks when the velocities are above $\sim22$ km s$^{-1}$.
For some regions the H$_2$ line fluxes are reproduced by a combination
of two C-shocks, in other regions the C-shocks (J-shocks) do not
sufficiently populate the highest (lowest) energy levels. In those
cases, we fit the H$_2$ line fluxes by a combination of C- and J-shocks.
The shocks travel through the clouds in $\sim10^4-10^5$ years (for 1 pc clouds), which is much shorter than the
dynamical time of the outflow ($10^8$ yrs - \citet{lehnert99}). Using the estimates of the warm
molecular gas mass and the typical cooling time of the shocked H$_2$
gas ($t_{cool}\sim10^4$ yr for $n_H = 10^3$ cm$^{-3}$), we can estimate
the mass flow of shocked gas, $M_{warm}/t_{cool}\sim7.5\times10^6
M_{\odot}/10^4$ yr $=750 M_{\odot}$ yr$^{-1}$, assuming that all the warm H$_2$ is shock-heated. The dynamical timescale of
the H$_2$-rich outflow is $\sim4.5$ kpc$/200$ km s$^{-1}=2.5\times10^7$
yr implying a total mass of shocked gas $\approx 5\times10^9 M_{\odot}$. This estimate is likely to be an upper limit since
the H$_2$ gas may not extend over the whole outflow and because this assumes
that the gas is only shocked once, which is highly unlikely.
The high mass flow rate of shocked gas may require the reformation of
H$_2$ in the post-shocked gas \citep[see][]{guillard09,guillard10}
to prolong the survival time of the H$_2$ gas in the outflow
\citep[e.g.][]{fragile04}.  Clouds may also survive longer if the
radiative cooling time is shorter than the time scale necessary for
the shocks to cross the cloud and the growth rate of Kelvin-Helmholtz
instabilities \citep{cooper08}. Such processes would lower the total
molecular mass required to explain the molecular emission from the
outflow, although for the C shocks considered here and even for many of the J shocks as
well, there is little destruction of H$_2$. At any rate, the total gas mass that is shock heated is similar to that of the cold molecular gas. Therefore the reservoir for the shocked gas is likely the cold molecular gas as probed through the CO emission from the wind.

\subsection{PAH destruction}

Since we see that the PAHs grains are on average larger and more ionized in the wind, it is natural to ask whether or not some of the enhanced H$_2/PAH$ could be do to a reduction in the PAH emission, rather than an enhancement of H$_2$.
The decrease in PAH flux is probably just due to a decrease of the radiation field, but there is also a change in the grain properties associated with the wind.
The gradient of decrease in H$_2$ emission is significantly lower than that of the PAH emission, suggesting an additional heating source for the H$_2$ (see Section 3.2). 
One of the ways PAH emission can decrease relative to H$_2$ is PAH
destruction. PAH grains can be destroyed either by hard radiation
field, X-rays, or shocks. The EWs of the PAH features in starburst
galaxies are observed to decrease sharply when [NeIII]$15.6\mu$m/[NeII]$12.8\mu$m$>2$, where
[NeIII]/[NeII] traces the hardness of the radiation field. This has been
interpreted as evidence of PAH destruction by intense radiation fields
\citep{wu06, beirao06}. In M82, despite an increase of [NeIII]$15.6\mu$m/[NeII]$12.8\mu$m
in the outflow, in most of the outflow [NeIII]$15.6\mu$m/[NeII]$12.8\mu$m remains less than 1.
[NeIII]$15.6\mu$m/[NeII]$12.8\mu$m$\sim 1$ indicates that there is hard ($> 40.96$ eV) radiation present, but not enough to have an effect on the average PAH properties. 
Furthermore, only a small fraction of the gas in the M82 outflow is photoionized, and PAHs may survive in HI and H$_2$ gas. 
PAH destruction is also very sensitive to
the clumping factor of the gas. If the gas is dense and clumpy with
relatively small volume filling factor and large covering fraction, the PAH grains will be shielded from the hard photons in the radiation field. So PAH destruction by hard photons is not likely to be the cause of the increase of H$_2$/PAH in the wind.

PAH grains are rapidly destroyed in hot, X-ray emitting gas
\citep{micel10b}. However, the presence of PAHs at large distances (10 kpc
from the disk) in the outflow shows that they are protected in cooler
phases of the gas, mostly in molecular gas. It is possible, however,
that X-rays from the hot outflow fluid could destroy the PAH carriers, but
the correlation between the X-ray emitting gas and the PAH maps, particularly in the Southern outflow, suggests that grain destruction by X-rays in the wind is not dominant.

Thermal collisions with very small and big grains can
have a significant impact on the cooling of the hot gas ejected into the
halo \citep{dwek90}. However, the impact of this cooling will generally
be short lived since the cooling efficiency drops as the grain sputtering
occurs in the hot gas \citep{guillard09}, unless the hot halo gas is
constantly replenished with dusty gas. A physical dust cooling model
is needed to fully assess whether thermal emission from collisionally
heated dust in the hot plasma is a significant contribution to the gas
cooling in the halo of M82, and whether this contribution changes as a
function of the radius.

Another possible mechanism for PAH destruction is shocks. As seen in
Figure~\ref{h2figshock}, the H$_2$ emission that we see can be fit with shock models having speeds of up to 40 km s$^{-1}$. PAH destruction likely only becomes relevant for shock
velocities $>$100 km s$^{-1}$ \citep{micel10a}. It is unlikely that the low velocity shocks that heat the molecular gas and enhace the H2 emission are also destroying the PAH grains, although we clearly see a change in the grain population. Similarly, \citet{contursi13} find that shocks do not dominate the [CII] and [OI] emission in the wind.
Overall, the enhancement in
H$_2/PAH$ is likely caused by H$_2$ excitation by shocks, although we
cannot rule out a relatively small effect due to PAH destruction.

In dense molecular clouds, where the PAH destruction by shocks is less efficient, 
cosmic rays are the most efficient source of destruction of PAH molecules, mostly
because of their bombardment by energetic ions \citep{micel11}. This is
likely to play an important role in the M82 wind, because the intense star
formation in M82 is likely to enhance the cosmic ray intensity by a factor of
$\approx 5$ compared to the Milky Way \citep{robitaille10}. Theoretical studies suggest that CRs can destroy PAHs in dense molecular clouds, but in these environments the heating will also be very inefficient, causing a reduction of PAH emission.
\citet{micel11} showed that only PAHs with a number of carbon atoms greater
than $\approx 250$ have a survival lifetime greater than the circulation time
scale of the outflow ($\approx 2 \times 10^8$~yr). Thus, there may be a
size-selection effect of PAHs in the outflow, which is roughly consistent with the
size inferred from the band ratios (we mostly see PAH $6.2/7.7< 0.3$, meaning an
average size greater than 250 carbon atoms, see Figure~\ref{drainefig}), 
and cosmic rays are likely to play a role. 

The calculations of Micelotta et al. 2010 show
that any PAH shocked at $> 100$ km/s should be destroyed. To be
re-formed in the post-shock gas, an efficient and fast re-formation
process is needed. PAHs are a product of high temperature chemistry
involving abundant carbon bearing precursors such as CH$_4$ and C$_2$H$_2$. In
the cooled postshock gas these precursor species are likely not very
abundant. In the cold neutral medium, the global PAH coagulation timescale is of the
order of 100 Myr, much longer than the outflow dynamical time \citep{seok14}. AKARI results from \citet{yamagishi12} suggest that very small grains still exist in the halo. Therefore it is reasonable to assume that PAH are produced in
the disk, carried up in the outflow, and "protected" from fast shocks
in the denser material. If molecular clouds are formed in
the postshock gas in the outflow, then there may be PAH coagulation
over timescales comparable to the outflow time (coagulation timescales
are of the order of a few Myr in molecular clouds), but it should be very small
fraction of the volume, contrary to what is observed.

In conclusion, the enhancement of the H$_2$/PAH is likely caused by H$_2$ shock excitation, although we cannot rule out a relatively small effect due to PAH destruction.

\section{Conclusion}
\label{conclusion}

\begin{itemize}
\item We have mapped the Northern and Southern superwind in the nearby starburst galaxy M82 using the $Spitzer$-IRS.  We find strong H$_2$, PAH and Ne line emission that extends over nearly 5 arcminutes above and below the plane of the galaxy, and have measured the variation of the H$_2/PAH$, [NeIII]/[NeII] and PAH feature ratios throughout the wind.

\item We estimate a total warm ($100-500$ K) molecular gas mass to be between $\sim5-17\times10^6 M_{\odot}$ and the total kinetic energy of the warm molecular outflow to be between $\approx6-20\times10^{53}$
erg assuming an outflow velocity $v_{outflow}$=160 km s$^{-1}$. The warm molecular
gas provides about 0.5-5\% of the kinetic energy -- comparable to that
observed in the atomic gas.

\item There are clear variations in the PAH band ratios throughout the wind.  The 6.2/7.7 and 11.3/7.7 micron PAH band flux ratios, when compared to models, suggest a higher fraction of larger and/or more ionized PAHs in the wind compared to the starburst disk, especially in the Northern part of the wind.

\item The warm molecular gas to PAH ratio (H$_2/PAH$) is enhanced in
the outflow by factors of 10-100 as compared to the starburst disk. This enhancement in the H$_2/PAH$ ratio does not seem to
follow the trends with the ionization of the atomic gas (as measured
with the [NeIII]/[NeII] line flux ratio) seen in the starburst. Our results indicate that UV and X-ray emission are probably not the dominant excitation mechanisms of H$_2$. Models suggest the observed values could be reproduced with
slow C- and J-shocks ($v < 40$ km s$^{-1}$) driven into dense molecular
gas entrained in the outflow. Using a dynamical timescale of the H$_2$-rich outflow of about 10$^7$ years results in an upper limit of 5$\times$10$^9$ M$_{\odot}$ of molecular gas heated and shocked over the lifetime of the wind. Given the rate at which the gas must be shocked, the reservoir for this shocked gas is the cold molecular gas as observed in CO. 

\item The emission from all of these phases is likely due to clouds entrained in and accelerated by the flow. They are being
shock heated by the hot plasma with high ram pressure generated by the
collective thermalization of stellar winds and supernova in the starburst
region of M82. These clouds are multiphase and the variation in properties are likely due to a range of shock velocities which depend on the density of the pre-shocked gas which depend on the density of the pre-shocked gas.

\end{itemize}

\clearpage
\begin{table*}
 \centering
 \begin{minipage}{140mm}
\caption{PAH band and ionic line fluxes (in units of $10^{-16}$ Wm$^{-2}$) and ratios. The regions for the extracted spectra are shown in Figure~\ref{pahmapfig} and described in the text.}
  \begin{tabular}{lcccccccccc}
    \hline
Region & 6.2$\mu$m & 7.7$\mu$m & 8.6$\mu$m & 11.3$\mu$m & 17$\mu$m & PAH $6.2-12.6\mu$m & 6.2/7.7 & 11.3/7.7 &  [NeII] & [NeIII]\\
    \hline
1	&7.12&	23.31&	4.19&	6.44&	3.52&	44.58 & 0.305  &0.151 & 3.62 & 1.36\\
2&	1.86&	3.31&	1.28&	2.56&	0.920&  9.93 & 0.562	 &0.278 & 2.60 & 0.966\\
3&	1.01&	2.58&	0.526&	1.12&	 0.751& 5.987 &   0.391 &	0.291 &1.27 & 0.550\\
4&	0.811&	4.76&	0.895&	2.65&	1.76&   10.876 & 0.170 &0.370 & 0.536 & 0.893 \\
5&	12.53&	40.41&	8.93&	13.86&	8.32&	84.05 &0.310 &0.206 &2.17 & 5.38\\
6&	7.30&	22.03&	6.04&	10.65&	7.94&	53.96 &0.331 &0.360 &2.79 & 4.18\\
7&	9.39&	29.26&	6.61&	11.43&	7.47&   64.16 &0.321 &	0.255 & 6.06& 3.44 \\
8&	349.6&	1167.7&	254.3&	397.7&	282.6&  2451.9 &0.299 &	0.242 &339.3 & 89.4\\
9&	741.9&	2783.7&	513.5&	668.9&	254.3&	4962.3 &0.261 &0.091 &568.0 & 83.1\\
10&	10.3&	37.6&	82.7&	18.3&	10.7&   159.6 &	0.274 &0.285 &5.43 &2.57 \\
11&	30.9&	105.7&	22.2&	37.0&	21.3&   217.1 & 0.292	 &0.202 &30.2 & 9.91\\
12&	8.22&	27.2&	7.55&	15.0& 	9.40&	67.37 & 0.302 &0.346 & 9.37& 4.17\\
    \hline
\label{pahtab}
  \end{tabular}
\end{minipage}
\end{table*}

\clearpage
\begin{table*}
  \centering
 \begin{minipage}{140mm}
\caption{H$_2$ line fluxes (in units of $10^{-17}$ Wm$^{-2}$) and ratios. "Total" denotes the sum of all emission in the maps.}
\begin{tabular}{lcccccc}
\hline 
Region & H$_2$ S(0) & H$_2$ S(1) & H$_2$ S(2) & H$_2$ S(3) & H$_2$ S(1) - S(3)\\
\hline
1	&	$<0.313$	&	4.13$\pm$0.12&		0.840$\pm$0.142	&	1.70$\pm$0.14 & 6.67\\
2	&	$-$			&		4.35$\pm$0.15&		0.784$\pm$0.173&		3.94$\pm$0.10 &  9.07\\
3&		$<0.169$&		2.81$\pm$0.07&		0.517$\pm$0.267&		1.68$\pm$0.23 & 5.01\\
4	&	$<1.45$&		6.68$\pm$0.14	&	1.74$\pm$0.16	&	3.17$\pm$0.14 & 11.6\\
5	&	$<0.991$&		14.3$\pm$0.2&		3.93$\pm$0.24&		8.74$\pm$0.41 & 27.0\\
6	&	$<2.81$&		9.68$\pm$0.26&		6.24$\pm$0.55&		2.05$\pm$0.18 &18.0 \\
7&		$<1.25$&		8.87$\pm$0.14&		2.37$\pm$0.14	&	4.79$\pm$0.20 & 16.0\\
8&		$<48.7$&		37.8$\pm$0.6	&	32.2$\pm$0.4	&	83.4$\pm$0.7 & 153.4\\
9	&	$-$		&			179.8$\pm$0.8	&		84.6$\pm$0.3&		108.4$\pm$0.3 & 372.8\\
10&		$<0.528$&		23.9$\pm$0.2	&	5.73$\pm$0.11	&	119.0$\pm$3.0 & 148.6\\
11&		$<5.51$&		31.7$\pm$0.2	&	10.4$\pm$0.1&		16.6$\pm$0.1 & 58.7\\
12	&	$<0.613$&		8.85$\pm$0.12&		2.50$\pm$0.12&		4.94$\pm$0.21 & 16.3\\
Total & $<253.9$	&	1171.5$\pm$3.2&	455.1$\pm$2.7	&1010.0$\pm$5.6	& 2636.6\\
\hline
\label{h2fluxtab}
\end{tabular}
\end{minipage}
\end{table*}

\begin{table*}
  \centering
 \begin{minipage}{140mm} 
\caption{H$_2$ Temperature Model Parameters. "Total" denotes the sum of all selected regions.}
\begin{tabular}{lccccccc}
\hline 
Regions & T(K) & Ortho/Para & $N(H_2)$ & $M(H_2)$ & $L(H_2)$ \\
& & $\times10^{19}$ cm$^{-2}$ & $\times10^4 M_{\odot}$ & $\times10^4 L_{\odot}$ \\
\hline
Region1 &210. $\pm$   18. & 2.883 & 2.12 & 4.28$\pm$1.47 &  1.43 \\
	& 1094. $\pm$  208. &  3.000 &  0.05 & 0.0107$\pm$0.0036 &   1.83 \\ 
Region3 &  240$\pm$54 & 2.945 & 0.838 & 12.9$\pm$10.1 &  7.91 \\ 
	&  819$\pm$126 &  3.000 &  0.007 &  0.113$\pm$0.087 &   7.90 \\
Region4 & 236.   $\pm$ 14. & 2.940 & 2.07 & 4.19$\pm$ 0.82 &   2.26\\
	& 831. $\pm$   124. & 3.000 & 0.0128 & 0.03$\pm$0.01 &   1.84\\
Region5     &206. $\pm$   17. & 2.8871 & 6.77 & 13.7$\pm$4.1 &  4.32\\
	& 549.  $\pm$  54. & 3.000 & 0.132 & 0.27$\pm$0.11 &   4.37 \\
Region6   & 166.  $\pm$   21. & 2.653 & 13.5 & 27.3$\pm$18.2 &   3.25\\
	& 593. $\pm$   56. &  3.000 & 0.086  &  0.152$\pm$0.052 &  3.30 \\
Region7    & 224. $\pm$   12.  &2.917 & 3.30 & 6.66$\pm$1.17 &  2.96\\
	& 657.  $\pm$  75. & 3.000  &0.0382 & 0.0771$\pm$0.0297 &   2.68 \\
Region8 &  324.$\pm$12. & 2.994 & 5.32 & 10.7$\pm$0.6 &  64.0 \\
	&  595.$\pm$13. & 3.000 & 0.80 & 1.6$\pm$0.2 &  3.6 \\
Region10 &  218.$\pm$6. & 2.913 & 9.5 & 19.2$\pm$1.7 &  33.0 \\
	 &  637.$\pm$62. & 3.000 & 0.091 & 0.18$\pm$0.06 &  5.2 \\
Region11 & 233.  $\pm$   3. & 2.934 & 11.4 & 23.0$\pm$0.81 &   11.8 \\
	& 623.  $\pm$  14. & 3.000& 0.02 & 0.39$\pm$0.03 &  10.1\\
Region12 & 241. $\pm$   10. & 2.946 &  2.61 &  5.28$\pm$0.666 &   3.17 \\
	& 816. $\pm$  157.  & 3.000 &  0.02 &0.042  $\pm$0.002 &   3.09 \\
Total & 272$\pm$2 &  2.976 & 1.39 & 476$\pm$7.66 &  463 \\
 	& 854$\pm$18 & 3.000 & 0.023 & 7.67$\pm$0.42 & 640 \\
\hline
\label{h2tab1}
\end{tabular}
 
\end{minipage}
\end{table*}

\begin{table*}
  \centering
 \begin{minipage}{140mm} 
\caption{H$_2$ Shock Model Parameters. "Total" denotes the sum of all selected regions.}
\begin{tabular}{lccccccc}
\hline 
Regions& $v_{shock}$  & $M_{flow}$ & $t_{cool}$  & $M(H_2)$ & Type\\
 & km s$^{-1}$ & $M_{\odot}$/yr & $\times10^3$ yr & $\times10^4 M_{\odot}$ &  \\
\hline
Region1   &   5.5  &  42$\pm$3.1  &  1.6  &  6.8$\pm$0.5 &C \\
          &  35.0  &  16$\pm$1.4  &  3.4  &  5.4$\pm$0.47 & J \\
Region3  &  5.5  &  210$\pm$35  &  1.6  &  34$\pm$5.6 & C \\  
         & 33.0  &  150$\pm$11 &    3.2  &  48$\pm$3.5 & J\\
Region4  &   6.0  &  48$\pm$6.1 &    1.6  &  7.6$\pm$0.96 & C\\   
         & 35.0  &  16$\pm$2.4  &  3.4  &  5.5$\pm$0.83 & J \\
Region5  &  5.0  &  200$\pm$12  &  1.6  &  33$\pm$2.0 & C\\    
         & 20.0  &  1.9$\pm$0.14  &  2.1 &  0.41$\pm$0.03 & C\\
Region6  &  4.0  &  1200$\pm$200  &  1.7  &  200$\pm$33 & C\\    
         & 21.5  &  1.3$\pm$0.31  &  2.1  &  0.27$\pm$0.06 & C \\
Region7  &   5.0  &  130$\pm$32   & 1.6  &  21$\pm$5.2 & C \\  
         & 21.5  &  0.97$\pm$0.14 &   2.1  &  0.2$\pm$0.029 & C \\
Region8  &   5.5  &  400$\pm$91  &  1.6  &  64$\pm$15 & C \\    
         & 21.5  &  17$\pm$4.4 & 2.1 &  3.6$\pm$0.93 & C \\
Region10 &  5.5  &  210$\pm$40  &  1.6  &  33$\pm$6.3 & C\\    
         & 21.5  &  1.5$\pm$0.24  &  2.1 &    0.31$\pm$0.05 & C \\
Region11 &  5.0  &  500$\pm$63  &  1.6  &  80$\pm$10 & C\\    
         & 18.0  &  6.2$\pm$0.26 &  2.1 &  1.3$\pm$0.05 & C\\
Region12  &   5.5  &  84$\pm$6.3  &  1.6  &  13$\pm$0.98 & C\\   
          & 22.0  &  33$\pm$1.4 &  2.2 &   7.5$\pm$0.32 & J\\
Total &   5.5  &  10000$\pm$2100  &  1.6  &  1700$\pm$360 & C\\    
        & 21.5  &  180$\pm$14  &  2.1 & 39$\pm$3.1 & C\\
\hline
\label{h2tab2}
\end{tabular}
 
\end{minipage}
\end{table*}

\end{document}